%% file: main_arxiv.tex
\documentclass[12pt]{article}

\usepackage{geometry}
\usepackage{enumerate}
\usepackage{latexsym,booktabs}
\usepackage{amsmath,amssymb}
\usepackage{graphicx}
\graphicspath{ {./images/} }
\usepackage[singlespacing]{setspace}
\usepackage{amsthm}
\usepackage{svg}
\usepackage{hyperref}
\usepackage{bm}
\usepackage{rotating}
\usepackage{todonotes}
\usepackage{tikz}
\usepackage[authoryear]{natbib}
\usepackage{mathtools}
\usepackage[ruled,linesnumbered]{algorithm2e}
\usepackage{amsmath, amssymb}
\usepackage{subcaption}

\usepackage{titlesec}
\newtheorem{Proposition}{Proposition}

\numberwithin{Theorem}{section}
\numberwithin{Definition}{section}
\numberwithin{Lemma}{section}
\numberwithin{Algorithm}{section}
\numberwithin{equation}{section}

\usepackage[utf8]{inputenc}
\usepackage[english]{babel}

\usepackage{graphicx}
\usepackage{natbib} %
\usepackage{url} %

\begin{document}

\def\spacingset#1{\renewcommand{\baselinestretch}%
{#1}\small\normalsize} \spacingset{1}

  \title{\bf An Efficient Likelihood Ratio Test for Online Changepoint Detection in the Presence of Autocorrelation}
  \author{
    Yuntang Fan, Paul Fearnhead, Idris A. Eckley, Gaetano Romano \hspace{.2cm}\\
    School of Mathematical Sciences, Lancaster University\\
    % and \\
    % Paul Fearnhead %\thanks{    The authors gratefully acknowledge the support of EPSRC and BT via grants \textit{UKRI2698, EP/Z531327/1}}\hspace{.2cm}
    % \\
    % School of Mathematical Sciences, Lancaster University\\
    % and \\
    % Idris A. Eckley %\thanks{The authors gratefully acknowledge the support of EPSRC and BT via grants \textit{UKRI2698, EP/Z531327/1}}\hspace{.2cm}
    % \\
    % School of Mathematical Sciences, Lancaster University\\
    % and \\
    % Gaetano Romano %\thanks{The authors gratefully acknowledge the support of EPSRC via grants \textit{UKRI2698}}
    % \\
    % School of Mathematical Sciences, Lancaster University
    }
    
  \maketitle

\begin{abstract}
Changepoint detection methods have seen considerable development in recent years, with online algorithms capable of identifying structural changes in streaming data in near real time. However, the majority of existing methods are designed under the assumption of IID\ observations, rendering them susceptible to either more false positives or longer detection delays when applied to data exhibiting temporal dependence, a common feature of many real-world data streams. In this article, we extend the generalised likelihood-ratio (GLR) statistic to autoregressive processes of order $p$, and adapt the \texttt{focus} algorithm to develop a computationally efficient online change detector. The resulting \texttt{AR($p$)-focus} algorithm achieves an average computational cost of $\mathcal{O}(\log n)$ per iteration, making it suitable for high-frequency data streams. Through simulation studies, the proposed approach is seen achieving greater detection power than IID-based tests when the underlying data exhibit temporal correlation. We further illustrate the practical utility of \texttt{AR($p$)-focus} through an application to a real-world telecommunications dataset.
\end{abstract}

\noindent%
\vfill

\newpage
\spacingset{1.8} %

\section{Introduction}
Detecting structural changes in time-series data - commonly known as changepoint detection - is a fundamental statistical problem that arises in applications across informatics and cyber-security (\citealt{tartakovsky2005nonparametric, jeske2018statistical, chen2019sequential}), machine learning (\citealt{guan2025keeping}), engineering (\citealt{henry2010fault, pouliezos2013real, zaliskyi2025algorithms}), bioinformatics (\citealt{willenbrock2005comparison, liehrmann2023diffsegr}), finance (\citealt{financeintro}), and climate records (\citealt{climateintro}). A changepoint marks an abrupt shift in the underlying data-generating property, such as a change in mean, variance, or slope. While classical offline changepoint methods assume full access to the entire dataset, many modern applications require decisions to be made as data arrive, seeking to detect a change as soon as possible. Several online changepoint detection algorithms have been developed to identify simple changes in streaming data. For univariate data, these include Bayesian approaches (\citealt{bayesianonline1}; \citealt{fearnhead2007line}; \citealt{bayesianonline2}), and frequentist methods based on recursively applying likelihood
ratio or similar tests (\citealt{frequenonline1}; \citealt{frequenonline2}; \citealt{frequenonline3}).

Many real-world data streams exhibit serial dependence, often well modelled by auto-regressive (AR) processes of order p. Few examples include applications related to telemetry and Internet of Things (IoT), where systems frequently display autocorrelation and slowly-varying trends (\citealt{austin2024detection, yang2024communication}). Additionally, in astronomical observations, background rates vary systematically across timescales, demanding careful modelling (\citealt{crupi2023searching}). Stock prices or returns in financial time series often show autocorrelation at short lags (\citealt{intro_financeapp}). In environmental data, air-quality indices or river flow measurements, frequently follow AR-like dynamics due to slowly changing meteorological or hydrological conditions (\citealt{intro_airapp}; \citealt{intro_riverapp}).

In addition to serial dependence, modern monitoring problems often involve large-scale or high-frequency data streams, where both statistical efficiency and computational scalability are essential. In these settings, changes and anomalies must often be identified as quickly as possible, so that timely corrective actions or policy changes can be implemented.

\begin{figure}
    \centering
    \includegraphics[width=\linewidth]{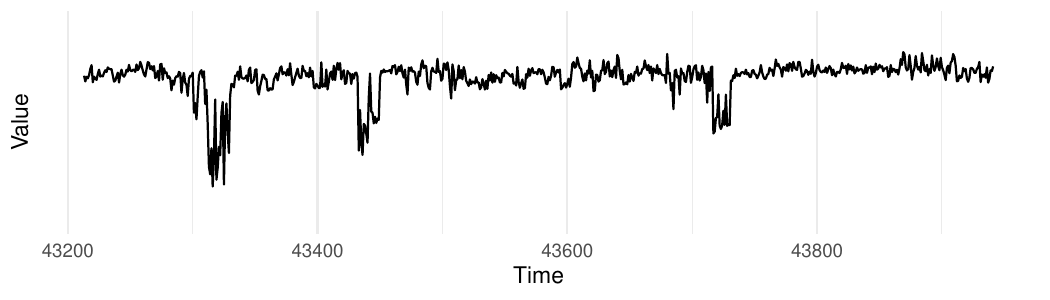}
    \caption{A subset of one time series from our network telemetry application.}
    \label{fig:intro_data}
\end{figure}

In particular, our work is motivated by a network monitoring application in which performance metrics are collected continuously from telecommunication network devices at high frequency. Under normal operating conditions these measurements exhibit strong temporal dependence. However faults, congestion events, or configuration changes can induce unexpected abrupt structural changes in the observed process (see, for example, Figure \ref{fig:intro_data}, where three sudden drops occur around times 43,320, 43,440 and 43,720). Effective monitoring therefore requires methods that can account for serial dependence while simultaneously providing fast online detection.

For temporally correlated observations, \citet{pre_ar_basseville1995detection} developed likelihood-based sequential detection procedures for mean and variance changes in AR processes. \citet{pre_ar_lai1995sequential} developed a general likelihood-based framework for sequential detection of mean and variance changes under dependent observations, which includes autoregressive time-series models as special cases. \citet{pre_ar_chu1996monitoring} proposed sequential monitoring procedures for detecting structural changes under dependent observations, explicitly allowing autoregressive error processes such as AR(p). \citet{pre_ar_gombay2008change} explicitly studied sequential detection of mean and variance changes in autoregressive AR(p) time series. Although these methods are likelihood-based and statistically optimal, their direct online implementation can be computationally demanding for long data streams, motivating the development of computationally efficient online algorithms that preserve exact likelihood ratio evaluation. 

The recent \texttt{focus} algorithm, introduced in \citet{FOCuSIID}, implements a fast and exact solution to the (generalised) likelihood-ratio test from \cite{Siegmund}. The most important contribution of the \texttt{focus} algorithm was to reduce the computational complexity of $\mathcal{O}(n)$ per-iteration of a direct calculation to an average of $\mathcal{O}(\log n)$ without any approximations, allowing for an implementation of the GLR test capable of dealing with a real-time data stream. The \texttt{focus} algorithm has been adapted to different models from the exponential family (\citealt{expFOCuS, PoissonFOCuS}), non-parametric models (\citealt{BinomialFOCuS}), and finally to multivariate observations \citep{mdFOCuS}. 

However, the \texttt{focus} algorithm and its extensions all assume an independent and identically distributed (IID) process. While it is possible to run the \texttt{focus} algorithm directly on data exhibiting strong temporal dependence, this would result in a loss of detection power. In this article, we address this limitation by deriving a new formulation of the GLR test tailored to autoregressive processes of order $p$. Specifically, the contribution of this work is twofold. We first derive the Generalised Likelihood Ratio (GLR) statistic for an AR($p$) process and adapt the computational technique of the \texttt{focus} algorithm to achieve efficient computation suitable for high-frequency data streams, that we call \texttt{AR($p$)-focus}. We then present empirical evaluations that (i) show the gains in detection power compared to IID\ tests on data exhibiting temporal correlation and (ii) demonstrate the practical utility of \texttt{AR($p$)-focus} on a real-data application drawn from a telecommunications setting. The theoretical properties of the likelihood-ratio test are not discussed here, as these rest on well-established foundations provided by \citet{Yulikeratio}, \citet{lordentheory}, and \citet{bayesianonline2}.

The remainder of this article is organised as follows. In Section \ref{sec:focus-intro}, we introduce the \texttt{focus} algorithm in the IID setting. Section \ref{method of AR.FOCUS} describes the extension to the AR($p$) case. In Section \ref{sim tests}, we present an empirical simulation study illustrating performances of the proposed approach. Finally in Section \ref{sec:real-application} we discuss an application to monitoring performances of network devices. Links to software and replication code are available at the end of this manuscript.

\section{Online change-in-mean with IID noise}\label{sec:focus-intro}

We begin by introducing the canonical model for detecting a change in mean. The observed data stream $\{x_t\}_{t = 1, \dots}$ are realisations of a process
\[
X_{t} = \mu^{(t)} + \epsilon_{t},   
\]
where $\{\epsilon_t\}_{t=1,2,\ldots}$ are IID Gaussian random variables with mean zero and known variance. We will assume throughout that the variance of the noise is known and, without loss of generality, we standardise the data so that the $\epsilon_t$ have unit variance. The mean parameter at time $t$, $\mu^{(t)}$, is assumed to undergo a single change at an unknown time $\tau$, such that $\mu^{(t)}=\mu_0$ for $t\leq \tau$ and $\mu^{(t)}=\mu_1$ for $t>\tau$, where $\mu_0$ and $\mu_1$ denote the pre-change and post-change means, respectively. To simplify exposition, we will initially assume the pre-change mean $\mu_0$ is known, and without loss of generality set $\mu_0=0$. The method can be extended to $\mu_0$ being unknown, but we delay describing this extension to Section \ref{sec:unknown pre-change mean}. 

At time $n$ we wish to detect whether there has been a change prior to $n$. Conditional on a putative changepoint time, $\tau<n$, and post-change mean, $\mu_1$, the log-likelihood ratio test statistic (which, henceforth, we will call the likelihood-ratio statistic) for a change at $\tau$ to mean $\mu_1$ is \cite[see e.g.][]{FOCuSIID}
\begin{equation} \label{eq:LR_IID}
LR_{\tau,n}(\mu_{1})= \sum_{i=\tau+1}^{n}\mu_{1}(2x_{i}-\mu_{1}).
\end{equation}
The corresponding test-statistic for a change prior to $n$ involves maximising this over $\tau$, and we would detect a change if
\[
LR_n=\max_{\tau=0,\ldots,n-1} \max_{\mu_1} LR_{\tau,n}(\mu_{1}) > \lambda_n,
\]
for some chosen threshold $\lambda_n$ which may depend on $n$. Often $\lambda_n$ is a constant chosen based on some required behaviour if there is no change, such as the expected detection delay (\citealt{lordentheory}), e.g. the expected time to detect a change if there is none. Alternatively, we can let $\lambda_n$ increase with $n$ and choose it so that we control the probability of ever detecting a change if there is none \cite[]{frequenonline3}.

As presented, calculating the test statistic, $LR_n$, at time $n$ involves a computational cost that increases with $n$ due to the maximisation over $\tau=0,\ldots,n-1$. In many online applications such a linearly increasing computational cost is impracticable. We can reduce this cost by restricting our maximisation to a small grid of possible $\tau$ values at each iteration (\citealt{MOSUMpaper_Eiauer}; \citealt{MOSUMpaper_chu}) but this introduces an approximation. Instead, \cite{FOCuSIID} showed that it is possible to calculate $LR_n$ in an online setting with a computational cost that only increases logarithmically with $n$ on average, which they called the \texttt{focus} algorithm.

The Gaussian change-in-mean \texttt{focus} begins by solving the recursion of \citet{page1954continuous, page1955test} for a known pre-change mean, $\mu_0=0$.  If we define $LR_n(\mu_1)=\max_{\tau\leq n}LR_{\tau,n}(\mu_1)$, the likelihood ratio statistic conditional only on the post-change mean, then for $n=2,3,\ldots$
\begin{align}\label{recursion form iid}
 LR_{n}(\mu_1) = \max\{0,LR_{n-1}(\mu_1)\} + \mu_1(2x_n-\mu_1).   
\end{align}

The idea of \texttt{focus} is to solve this recursion jointly for all $\mu_1$. That is, at any time $n$ we solve for $LR_n(\mu_1)$, a function of $\mu_1$. This function is piecewise quadratic, and at each iteration the algorithm maps a piecewise quadratic and a new data point, to an updated piecewise quadratic, e.g.  $LR_{n-1}(\mu_1), y_n  \rightarrow LR_n(\mu_1)$. This is achieved with a cost that is on average $O(1)$ per iteration. Then to calculate $LR_n$ we just maximise this piecewise quadratic over $\mu_1$. In practice, the piecewise quadratic can be written as the maximum of a set of quadratics, each one corresponding to one $LR_{\tau, n}(\mu_1)$. Thus, we can find its maximum by maximising each quadratic individually and then finding the maximum of these maxima. Now, at iteration $n$ we would need to store and check $n$ separate curves, but in reality not all curves contribute to the final cost as for all $\mu \in \mathbb{R}$, the set of $LR_{\tau, n}(\mu_1) = LR_n(\mu_1)$ at time $n$ is bounded in expectation by $\log n$. Hence it is possible to drop (or prune) those functions that do not contribute (and never will) to $LR_n(\mu_1)$, and therefore the per-iteration cost of the \texttt{focus} algorithm in calculating $LR_n$ increases like $\log n$. 

For our use of the \texttt{focus} algorithm below it is helpful to briefly mention some features of the algorithm. At each iteration, $n$, the list of quadratics increases by one, as we add a quadratic that corresponds to the likelihood ratio test statistic given a change at time $n-1$. The algorithm then prunes quadratics that correspond to the possible change-point times that can never be optimal in the future regardless of the future data. The key to the computational efficiency of \texttt{focus} is that we can prune quadratics in the order of how recent their corresponding change-point is, such that once we do not prune a quadratic we know that all remaining quadratics would also not be pruned. This means that on average the number of quadratics we need to check for pruning is $O(1)$.

If we wish to estimate where a detected change is, we can use the estimator
\[
\hat{\tau} = \arg\max_{\tau} \left\{\max_{\mu_1} LR_{\tau,n}(\mu_{1})\right\}.
\]
The estimator $\hat{\tau}$ corresponds to the change-point time associated with the quadratic stored by \texttt{focus} that has the largest maximum value, and is trivial to obtain.

\section{\texttt{focus} in the presence of auto-correlation}\label{method of AR.FOCUS}

In many applications where we wish to detect a change in mean, it is inappropriate to assume that the noise is independent across time. We will show how the \texttt{focus} algorithm can be extended to the case where the noise is modelled as an $AR(p)$ process, leading to an algorithm that we call \texttt{AR($p$)-focus}. 

Our model for the data is now
\begin{equation} \label{eq:AR model}
    X_t=\mu^{(t)}+Z_t,
\end{equation}
where as above $\mu^{(t)}=\mu_0$ for $t\leq \tau$ and $\mu^{(t)}=\mu_1$ for $t>\tau$, but now we model $Z_t$ as an $AR(p)$ process. That is
\[
Z_t=\rho_{1} Z_{t-1} + \rho_{2}Z_{t-2} + \cdots + \rho_{p}Z_{t-p} + \epsilon_{t},
\]
with, $\epsilon_t$ being IID Gaussian random variables with mean 0. We initially assume $\mu_0$ is known as are $p$, $\rho_1,\ldots,\rho_p$ and the variance of the $\epsilon_t$. To simplify notation, we standardise the data so $\mu_0=0$ and each $\epsilon_t$ has unit variance. The case where $\mu_0$ is unknown is covered in the next section. We then discuss how to estimate the parameters of the $AR(p)$ process and how to implement \texttt{AR($p$)-focus} in a way that accounts for estimation error in Section \ref{tuning threshold}.  

Assume we have data $x_{-p+1},\ldots,x_1,x_2,\ldots$ that are realisations of (\ref{eq:AR model}), and at any time $n\geq 1$ wish to detect if a change has happened at time $\tau\geq0$. It is helpful to work with a whitened version of the data, so for $t=1,2,\ldots,$ define
\[
y_t= x_t -\sum_{i=1}^p \rho_{i}x_{t-i}.
\]
It is simple to show that for $t\leq \tau$, $y_t=\epsilon_t$, $y_{\tau+1}=\mu_1+\epsilon_{\tau+1}$ and for $t>\tau+1$,
\[
y_t= \left(1-\sum_{i=1}^{p \wedge (t-\tau-1)} \rho_i\right)\mu_1+\epsilon_t,
\]
where $p \wedge (t-\tau-1)$ is the minimum of $p$ and $t-\tau-1$. If we define $v_1=1$, $v_j=1-\sum_{i=1}^{j-1} \rho_i$ for $j=2,\ldots,p$, and $v_j=1-\sum_{i=1}^p \rho_i$ for $j>p$, then this can be simplified to, if $t=\tau+i$,
\[
y_t=v_i\mu_1+\epsilon_t, \mbox{ for $i=1,2\ldots$.}
\]

As in Section \ref{sec:focus-intro}, to detect a change prior to time $n$ based on data $x_{-p+1},\ldots,x_n$ we will use the likelihood-ratio statistic. If we condition on a changepoint time $\tau\in \{0,\ldots,n-1\}$ and post-change mean $\mu_1$ then the likelihood-ratio statistic is
\begin{equation}
\label{eq:LR-known} 
LR_{\tau,n}(\mu_1) =\sum_{t=\tau+1}^n v_{t-\tau} \mu_1(2y_t-v_{t-\tau}\mu_1).    
\end{equation}

This is equivalent to (\ref{eq:LR_IID}) applied to data $y_1,\ldots,y_n$ but allowing for the varying mean of $y_t$ after the change at $\tau$.

For $t>\tau+p$, the post-change mean is constant, and thus we can directly apply the idea of \texttt{focus} to give us an efficient approach to calculate the likelihood-ratio statistic. We now derive such an algorithm, which is more complex that the standard FOCuS algorithm as we need to separately treat the case of change within $p$ times-steps of the curent time and one which is older. To do this it is helpful to define the contribution to the likelihood-ratio statistic of a change at different lags from the current time.

Define the contribution of data $y_t$ to the likelihood-ratio statistic for a change at $\tau=t-k$, for $k=1,\ldots,p$ as
\[
C_{k,t}(\mu_1) = v_k\mu_1(2y_t-v_k\mu_1),
\]
and the contribution to the likelihood-ratio statistic for a change at $\tau<t-p$ as
\[
C_{t}(\mu_1) = v_{p+1}\mu_1(2y_t-v_{p+1}\mu_1).
\]

We now define the likelihood-ratio test statistic for a change within $p$ time-steps of the current time.
If we fix a change at time $\tau$ and consider the likelihood-ratio statistic given data $y_1,\ldots,y_{t}$ for $t\in\{\tau+1,\ldots,\tau+p\}$, then this can be shown to be 
\[
S_{t,\tau}(\mu_1):= \sum_{i=1}^{t-\tau} C_{i,\tau+i}(\mu_1).
\]

Finally, we need to also deal with changes that have occurred further in the past. The following proposition does so and gives a recursion for calculating $LR_n(\mu_1)$, the likelihood-ratio statistic conditional on $\mu_1$ but where we have maximised over $\tau$.
\begin{Proposition} \label{Prop:1}
For $n\geq p+1$, 
\begin{equation}\label{eq:LR AR}
LR_n(\mu_1)=\max\left\{ ~~\max_{j=1,\ldots,p} \{S_{n,n-j}(\mu_1)\}~~ , ~~ Q_n(\mu_1) ~~ 
\right\},
\end{equation}
where $Q_n(\mu_1)$ satisfies the recursion 
\[
Q_{p+1}(\mu_{1})= S_{p,0}(\mu_1)+C_{p+1}(\mu_1),
\]
and for $n=p+2,\ldots,$
\begin{equation} \label{eq:recursion AR}
    Q_n(\mu_1) = \max\left\{
S_{n-1,n-p-1}(\mu_1),Q_{n-1}(\mu_1) \right\} + C_{n}(\mu_1).
\end{equation}
For $n\leq p$, $LR_n(\mu_1) = \max_{i=1,\ldots,n} S_{n,n-i}(\mu_1)$.
\end{Proposition}

The idea of \texttt{AR($p$)-focus} is that we can use the \texttt{focus} algorithm to solve for $Q_n(\mu_1)$. To calculate the likelihood-ratio statistic at time $n$ we then maximimise $Q_n(\mu_1)$ and also maximise each of the $S_{n,n-j}(\mu_1)$ functions that appear in (\ref{eq:LR AR}), and take the maximum of these maxima. As for the IID case we can also obtain the estimated change-point location as the time corresponding to the quadratic, whether it is one of the quadratics that defines $Q_n(\mu_1)$, or is one of the $S_{n,n-j}(\mu_1)$s, that gives this maximum.

\subsection{Unknown pre-change mean} \label{sec:unknown pre-change mean}

We can extend our $AR(p)$ approach to deal with the pre-change mean being unknown. How we do this follows closely the approach for the IID case \cite[see][]{FOCuSIID}. The likelihood-ratio test statistic for a change at $\tau$, from pre-change mean $\mu_0$ to post-change mean $\mu_1$, can be written as
\begin{equation} \label{eq:LR-unknown}
LR^*_{\tau,n}(\mu_0,\mu_1) = 2\left( \ell_{\tau,n}(\mu_0,\mu_1) - \max_{\mu} \ell_{\tau,n}(\mu,\mu)\right),
\end{equation}
where
\[
\ell_{\tau,n}(\mu_0,\mu_1) =-\frac{1}{2}\sum_{t=1}^\tau (y_t-v_{p+1}\mu_0)^2 -\frac{1}{2} \sum_{t=\tau+1}^n (y_t-v_{t-\tau}(\mu_1-\mu_0)-v_{p+1}\mu_0)^2,
\]
is, up to an additive constant, the log-likelihood for a model with pre-change mean $\mu_0$ and post-change mean $\mu_1$ with a change at $\tau<n$. Here $y_t=x_t-\sum_{i=1}^p \rho_ix_{t-i}$, $v_1=1$ and $v_j=1-\sum_{i=1}^{(j-1)\wedge p}\rho_j$ as in the previous section.

Let $\mathcal{T}_n$ be the set of $\tau$ values such that if for some $\mu_0$ and $\mu_1$, $\ell_{\tau',n}(\mu_0,\mu_1)=\max_{\tau} \ell_{\tau,n}(\mu_0,\mu_1)$, then $\tau\in \mathcal{T}_n$. Then we can calculate the maximum of (\ref{eq:LR-unknown}) using 
\[
\max_{\tau,\mu_0,\mu1} LR^*_{\tau,n}(\mu_0,\mu_1) = 2\left(
\max_{\tau\in\mathcal{T}_n} \left\{\max_{\mu_0,\mu_1} \ell_{\tau,n}(\mu_0,\mu_1)\right\} - \max_{\mu} \ell_{\tau,n}(\mu,\mu)\right).
\]
This expression uses the fact that $\max_{\mu} \ell_{\tau,n}(\mu,\mu)$ is the same for all $\tau$.

We will define the set $\mathcal{T}_n$ to always include changes within $p$ of the current time $n$, and use the following result to keep only a subset of changes prior to $n-p$.

\begin{Proposition} \label{Prop:unknown mean}
At time $n$, consider two changepoint locations $\tau$ and $\tau'$ with $n-p>\tau'>\tau>0$. Let $h=\tau'-\tau$. There exists constant $b_{\tau,\tau'}$, that depends on the data $y_{\tau+1:\tau'+p}$, such that
\[
\ell_{\tau,n}(\mu_0,\mu_1)-\ell_{\tau',n}(\mu_0,\mu_1) > 0,
\]
if and only if $\mu_1>\mu_0$ and $\mu_1< b_{\tau,\tau'}-\mu_0$; or $\mu_1<\mu_0$ and $\mu_1>b_{\tau,\tau'}-\mu_0$.
\end{Proposition}
 The proof is given in Appendix \ref{Appendix: proof unknown mean}.  

If we fix $\mu_0$ then the candidate $\tau$ values for which the log-likelihood for some $\mu_1$ is largest can be obtained by finding the interval, if any, where the log-likelihood for a change at $\tau$ is greater than the log-likelihood at $\tau'$ for all $\tau'\neq \tau$. This interval will be defined in terms of the intersection of the intervals where the log-likelihood at $\tau$ is greater than each of the log-likelihoods at $\tau'$. If we consider a positive change, say, then the log-likelihood at $\tau$ will be greatest for $\mu_1$ that satisfy
\[
\mu_1-\mu_0>\max\left\{0,~~\max_{\tau'<\tau} b_{\tau',\tau}-2\mu_0\right\}
\mbox{ and }
\mu_1-\mu_0 < \min_{\tau'>\tau} b_{\tau,\tau'}-2\mu_0.
\]
Thus, $\tau$ is a candidate for this value of $\mu_0$ if and only if
\[
\max\left\{0,~~\max_{\tau'<\tau} b_{\tau',\tau}-2\mu_0\right\} < \min_{\tau'>\tau} b_{\tau,\tau'}-2\mu_0 \Rightarrow 
\max\left\{2\mu_0,~~\max_{\tau'<\tau} b_{\tau',\tau}\right\} < \min_{\tau'>\tau} b_{\tau,\tau'}
\]
It is immediate that $\tau$ is a candidate for some $\mu_0$ value if and only if
\[
\max_{\tau'<\tau} b_{\tau',\tau} < \min_{\tau'>\tau} b_{\tau,\tau'}.
\]
This condition corresponds to the pruning condition for the pre-change mean known case in the limit as $\mu_0\rightarrow -\infty$, and moreover can be implemented efficiently as described in \cite{FOCuSIID}. Specifically, the pseudo-code for implementing this can be found in Appendix \ref{pseudo-code unknown pre prune}. In practice, the maximisation steps are implemented by storing and updating coefficients of each $LR^*_{\tau,n}(\mu_{0},\mu_{1})$, the complete details can be found in Appendix \ref{coef updateing unknown pre}.

\subsection{Estimating Parameters and Calculating the Detection Threshold}\label{tuning threshold}

The proposed method requires knowledge of the noise variance, the AR parameters, and choice of the detection threshold $\lambda_n$. In most real-world applications, the AR structure is not known \textit{a priori}. As is common in the online change-point setting \cite[]{pre_ar_chu1996monitoring,frequenonline2}, we estimate the AR parameters and noise variance from a probationary period with no change, that is, a sequence of observations drawn from historical data, or a short initial segment of the data stream during which no detection is performed. In the case of this article, in simulation studies and application, for a fixed AR order we estimate the autocorrelation parameters and noise variance from the training set using the Yule--Walker equations. If we also estimate the AR order we fit AR($p$) models for $p=1,\ldots,p_{\max}$, for some suitably chosen maximum order $p_{\max}$, and then estimate $p$ using AIC.

Whilst it is possible to derive closed-form expressions for the detection threshold, they tend to be conservative in practice. Instead, we suggest using the common practice of adopting a Monte Carlo calibration procedure \citep{FOCuSIID, chen2022high, mdFOCuS}. There are various statistical criteria that can be used to define the detection threshold.  For our empirical studies, we  control the false positive rate at a pre-specified level $\alpha$, for a fixed run length $N$. Though, the Monte Carlo approach we adopt can also be used for other criteria, for example see \cite{chen2022high} for an equivalent approach for setting a threshold to have a required average run length.

We will describe our procedure for the case where the AR($p$) model is unknown and we are using a probationary period of length $N_0$ to estimate them. We first estimate the AR($p$) model using the probationary data. We call a model with constant mean and with this choice of AR($p$) model the null model in the following.
We then implement the following steps $B$ times: (i) Simulate an independent datasets of the length $N+N_0$ under the null model; (ii) Taking the first $N_0$ data-points as a probationary period, we estimate the AR($p$) model parameters using the data from the probationary period; (iii) We run \texttt{AR($p$)-FOCuS} (or another sequential algorithms) on the last $N$ data-points, using the estimated AR($p$) parameters, and record the maximum of the trace of the test statistic. We then set the threshold $\lambda$ as the upper $\alpha$-quantile of the resulting empirical distribution of maximum statistics across the $B$ replicates. 

Importantly, this procedure accounts for the uncertainty in the estimates of the AR model -- in that each time when we run \texttt{AR($p$)-FOCuS} say, we run it with estimated parameters based on simulated probationary data of the correct length. If we know the AR model we can omit step (ii) and use the true parameter values in step (iii).

\section{Simulation Study}\label{sim tests}

In this section, we aim to evaluate performances of \texttt{AR($p$)-focus}.  We compare three methodologies: (i) \texttt{focus}, which applies the IID\ algorithm described in Section \ref{sec:focus-intro} directly to the raw data, ignoring any autocorrelation in the noise; (ii) \texttt{focus\_prewhiten}, which pre-whitens the data using either true or estimated AR parameters and then applies \texttt{focus} to the residual; and (iii) \texttt{AR($p$)-focus}, our proposed approach, incorporating either known or estimated AR parameters into the detection statistic. In all cases we use the same methods for estimating unknown parameters and setting the detection threshold, described in Section \ref{tuning threshold}. We present results for using the unknown pre-change mean version of all algorithms, with results for the known pre-change in Appendix~\ref{appendix:known mean}.

Unless stated otherwise, all experiments use the following default configuration: data length $n = 10000$, true changepoint at observation $5000$ and change of size $\delta = 5$. In all scenarios the noise is generated from AR($p$) processes with the following specifications: AR(1) with coefficient $\rho_1 \in \{0.2, 0.5, 0.8, 0.95\}$; AR(2) with coefficients $(0.7, 0.2)$; and AR(3) with coefficients $(0.7, 0.3, -0.1)$. 

The threshold was set to achieve false positive rate of $5\%$ for a sequence of length $n = 10000$. As a measure for detection power, we will present results in terms of detection delay, e.g. the difference from the stopping time (first time where each statistics passes the threshold) and true changepoint. Across the study, each set of results uses $100$ replications, with separate datasets used both for threshold calibration, parameter tuning (where present) and finally for evaluating detection performance.

\subsection{Performance under Known AR parameters}\label{oracle setting performance compare}

We begin by examining the idealised case in which the AR parameters are assumed known, so that all three methods operate without any estimation error. This setting allows us to isolate the effect of autocorrelation itself on detection performance, independent of any uncertainty introduced by parameter estimation.

\begin{figure}[!ht]
    \centering
    \begin{subfigure}{\textwidth}
        \centering
        \includegraphics[width=1.05\textwidth]{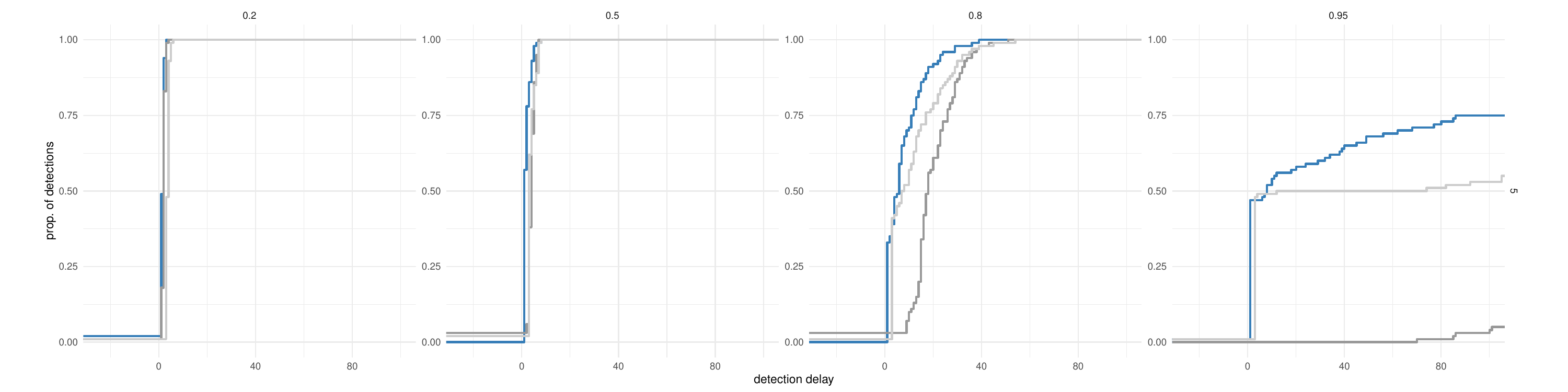}
        \caption{AR(1) 
        noise processes}
        
    \end{subfigure}

    \begin{subfigure}{\textwidth}
        \centering
        \includegraphics[width=.5\textwidth]{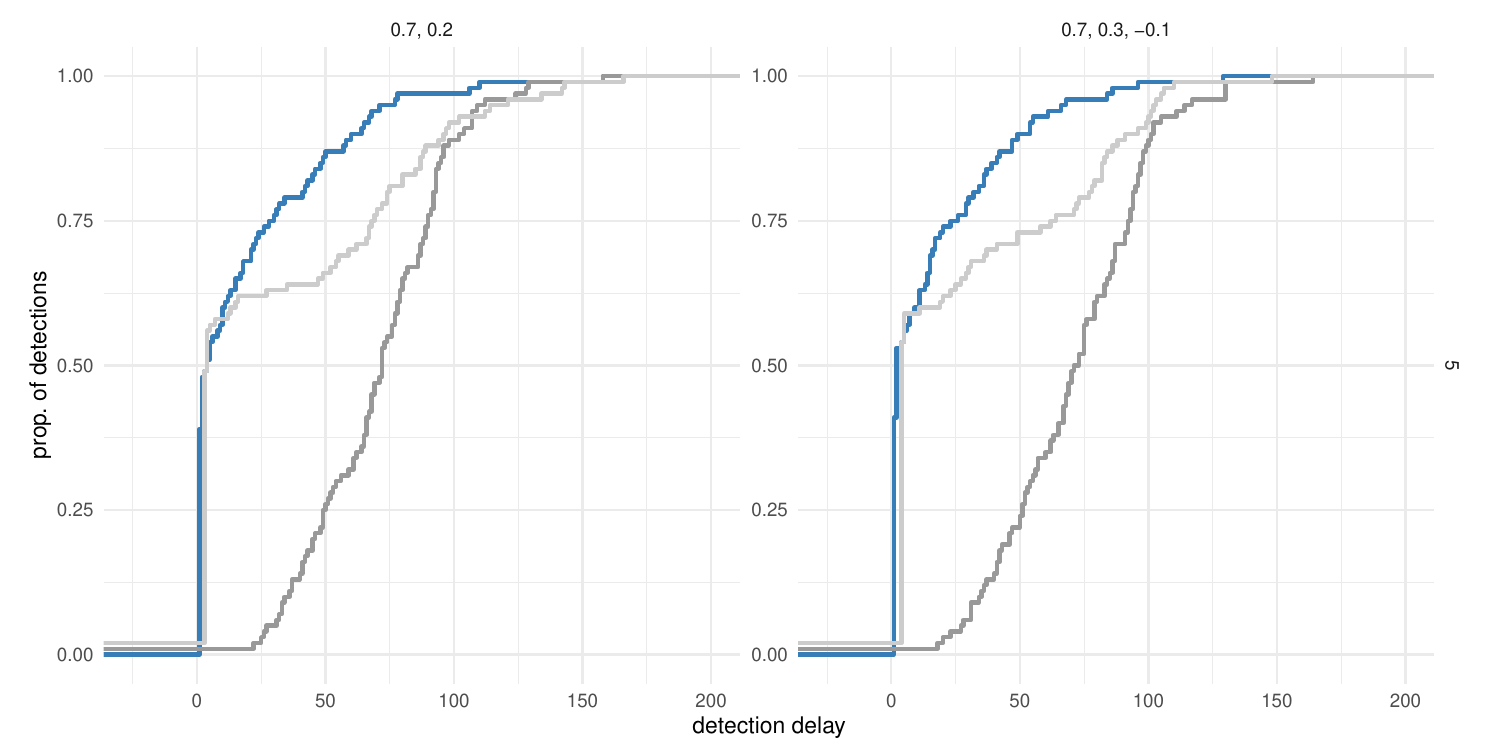}
        \caption{AR(2), AR(3) 
        noise processes}
        
    \end{subfigure}
    
    \caption{Oracle setting: empirical distribution of detection delay across different noise processes. Each column corresponds to a different AR(1) autocorrelation coefficient. Blue: \texttt{AR($p$)-focus}; light gray: \texttt{focus\_prewhiten}; gray: \texttt{focus}.}
    \label{fig:oracle_setting_comparison}
    
\end{figure}

Figure \ref{fig:oracle_setting_comparison} shows the empirical distribution of detection delay for each method and scenario (each plot reports the proportion of detection within a certain time delay). When autocorrelation is weak, all three methods perform comparably. As the strength of autocorrelation increases, however, the differences become more pronounced: \texttt{AR($p$)-focus} increasingly outperforms both \texttt{focus} and \texttt{focus\_prewhiten}, since the latter two methods fail to account for the dependence structure of the noise. This pattern holds across all AR(1), AR(2), and AR(3) processes, confirming that explicitly modelling the autocorrelation structure yields meaningful gains in detection efficiency when the noise is strongly correlated.

\begin{figure}
    \centering
    \includegraphics[width=0.5\linewidth]{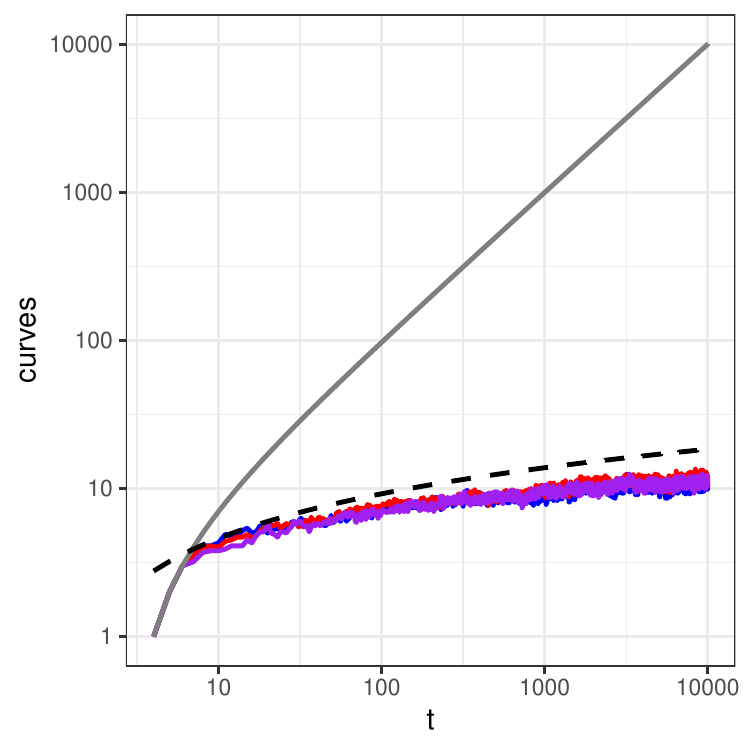}
    \caption{Oracle setting: number of curves stored at each time point by the brute-force method and AR($p$)-FOCuS. Purple: \texttt{AR($1$)} with 0.95; red: \texttt{AR($1$)} with 0.5; blue: \texttt{AR($1$)} with 0.25; gray: brute force method; black dashed: $2\log t$. Log-scale on both axes.}
    \label{fig:curves stored}
\end{figure}

To see the benefit of using \texttt{AR($p$)-focus} over a method that directly calculates the generalised LR statistic without pruning, we show in Figure \ref{fig:curves stored} the average number of quadratics stored by \texttt{AR($p$)-focus} for data simulated with no change under different AR(1) models. We see that the average number of quadratics stored at time $t$, and hence the average computational cost per iteration, is close to $\log t$, and varies little across different choices of the AR parameters.

\subsection{Performance under AR Parameter Estimation}\label{sec:sim_estimation}

We now estimate the AR($p$) parameters of the process. In this section parameters that are not fixed are estimated from a dedicated, separate training sequence, which we refer to as the probation period. To compare performances with the optimal setting, we add the \texttt{AR($p$)-focus} method where all parameters and order are assumed (denoted by ``oracle"), which corresponds to having an infinite length probation period.

\subsubsection{Effects of Probation Period}\label{training case performance compare}

Again, to remove nuisance parameters, we begin first by evaluating how probation period size affect the detection delay, assuming that the order $p$ of the AR($p$) process is known and estimating the coefficients alone. We vary the length of the probation period to assess how the amount of training data affects detection performance.

\begin{figure}[!ht]
    \centering
    \begin{subfigure}{\textwidth}
        \centering
        \includegraphics[width=\textwidth]{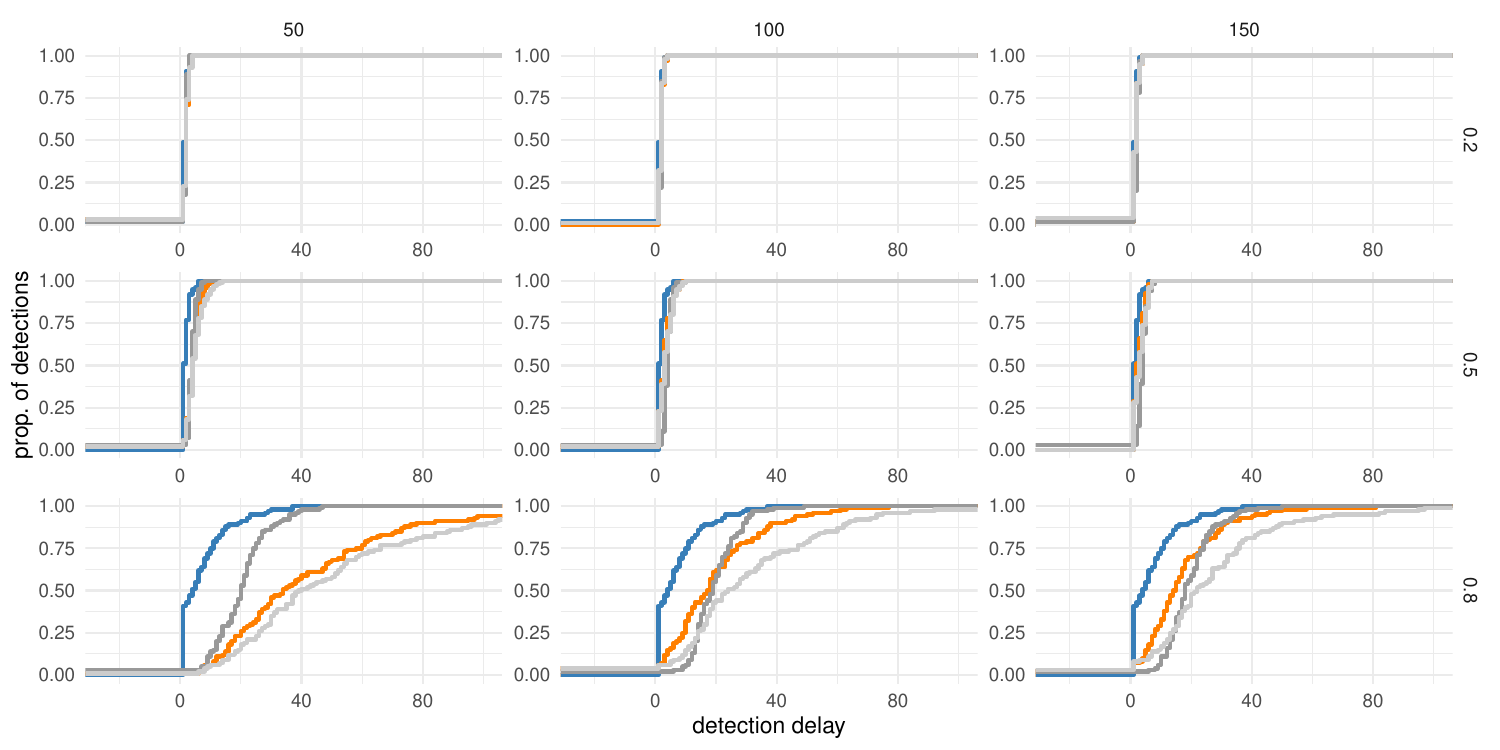}
        \caption{AR(1) noise processes}
        
    \end{subfigure}

    \vspace{1em}

    \begin{subfigure}{\textwidth}
        \centering
        \includegraphics[width=\textwidth]{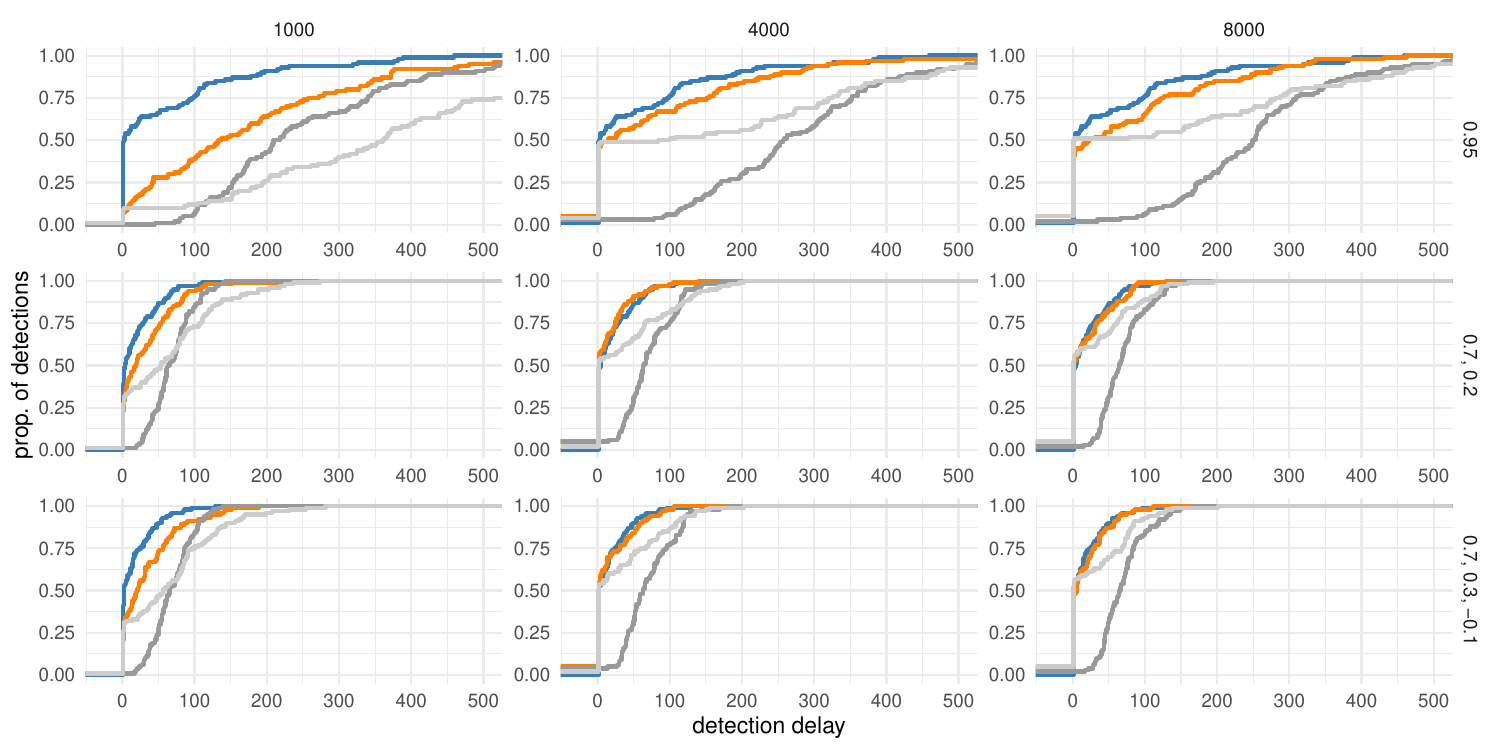}
        \caption{AR(1), AR(2), and AR(3) noise processes}
        \label{fig:probation_size_effect_ar3}
    \end{subfigure}

    \caption{Effect of probation period length on detection delay across different 
    noise processes. Each column corresponds to a different training size; each row 
    to a different AR process. Blue: \texttt{AR($p$)-focus} oracle; orange: \texttt{AR($p$)-focus}; gray: \texttt{focus}; light gray: \texttt{focus\_prewhiten}.}
    \label{fig:probation_size_effect}

\end{figure}

Figure \ref{fig:probation_size_effect} displays the detection delay distributions across a range of probation period sizes, for both weakly and strongly autocorrelated noise. Several patterns emerge clearly. First, detection is substantially easier when the underlying autocorrelation is weak, regardless of the method used, as in the previous section. Second, when the probation period is short and autocorrelation is strong, the performance of \texttt{AR($p$)-focus} degrades noticeably, falling below that of the simpler \texttt{focus} method: the additional parameters introduced by the AR model cannot be estimated reliably from limited data. However, as the probation period grows, \texttt{AR($p$)-focus} progressively recovers and eventually surpasses both \texttt{focus} and \texttt{focus\_prewhiten}, converging towards its oracle performance. These results highlight a fundamental trade-off: the benefits of modelling the autocorrelation structure are only realised once the probation period is sufficiently long to support accurate parameter estimation.

\subsubsection{Robustness to AR Order Misspecification and Unknown Order}\label{robustness order p}

The experiments in Section \ref{training case performance compare} assumed that the AR order $p$ is known. In practice, however, the order must also be determined from the data. We examine this in two stages: first assessing robustness to order misspecification, and then considering a fully data-driven procedure in which both the order and the AR parameters are estimated from the probation period.

\begin{figure}[ht]
    \centering
    \includegraphics[width=1\textwidth]{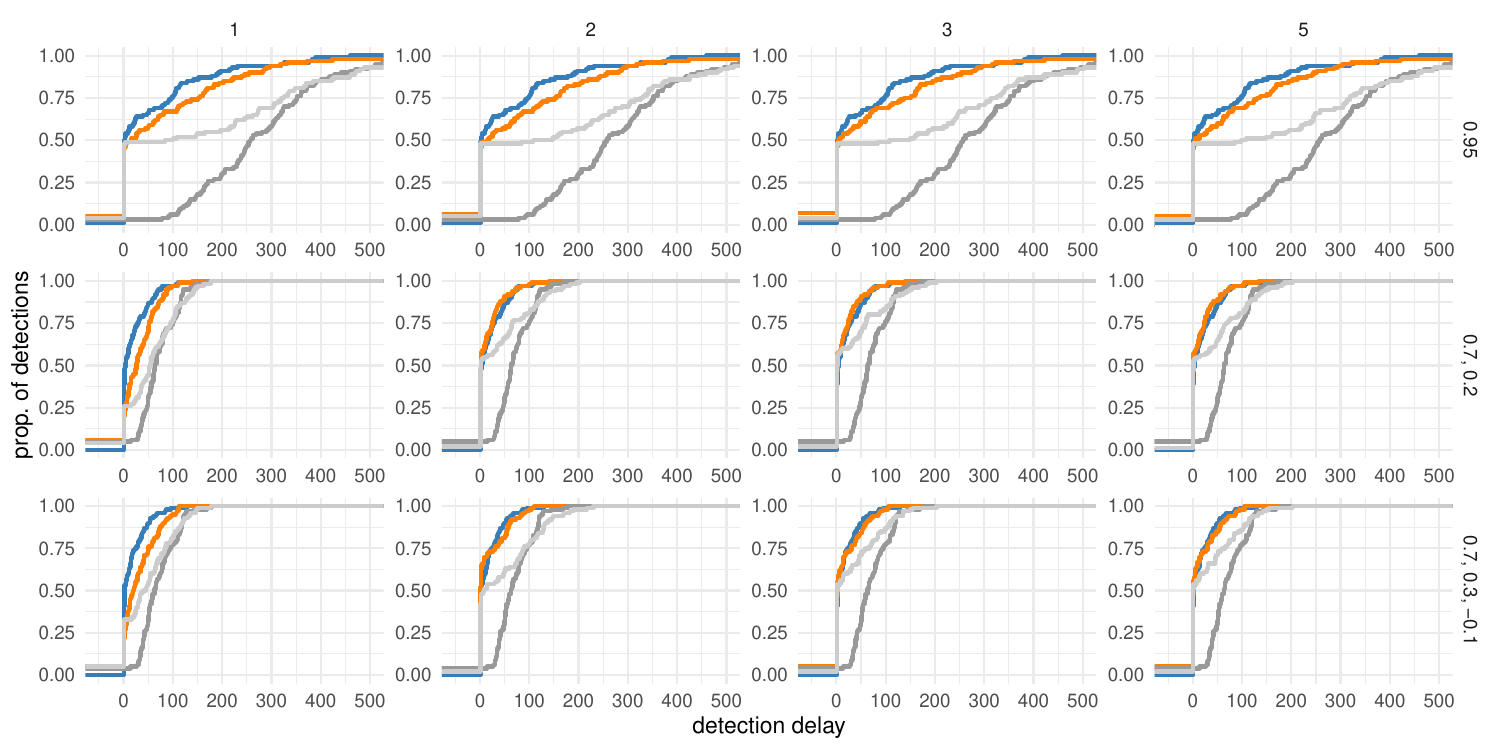}
    \caption{Robustness to AR order misspecification. Each row corresponds to a different true AR process (labelled by autocorrelation); each column to a different assumed order $p$ used by the algorithm. Blue: \texttt{AR($p$)-focus} oracle; orange: \texttt{AR($p$)-focus}; gray: \texttt{focus}; light gray: \texttt{focus\_prewhiten}.}
    \label{fig:Robustness order p test}
\end{figure}

To assess robustness, for each scenario, for a probation period size of $4000$, the AR order used when running the algorithm is fixed and varied over $\{1, 2, 3, 5\}$, covering cases of both under- and over-specification. Figure \ref{fig:Robustness order p test} summarises the results. When the assumed order is at least as large as the true order, both \texttt{AR($p$)-focus} and \texttt{focus\_prewhiten} incur only a modest loss of power: the additional parameters estimated are unnecessary but do not substantially distort inference. Conversely, when the assumed order is smaller than the true order, performance deteriorates more sharply, as the fitted model fails to capture the full autocorrelation structure of the noise, suggesting that, when in doubt, it is preferable to over-specify the order rather than under-specify the order.

An additional illustration of how detection performance varies with the strength of the AR(1) parameter for different assumed maximum orders is provided in Appendix \ref{appendix:max order p}, where we restrict attention to the AR(1) case for clarity.

\begin{figure}[ht]
    \centering
    \includegraphics[width=1\textwidth]{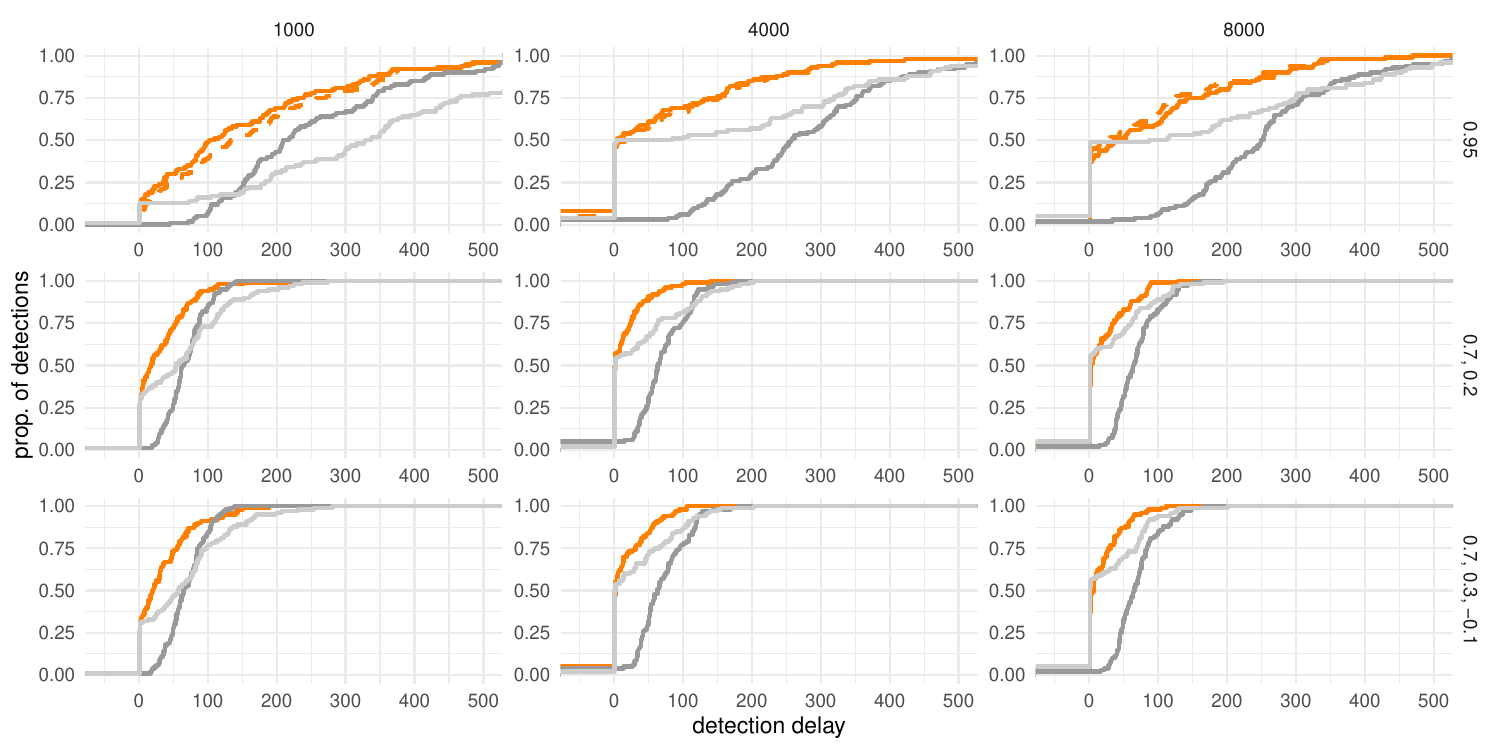}
    \caption{Performance of \texttt{AR($p$)-focus} when the AR order is unknown and selected by AIC. Each row corresponds to a different AR process (labelled by autocorrelation); each column to a different probation period length. Orange dashed: \texttt{AR($p$)-focus} (known p); orange: \texttt{AR($p$)-focus}; gray: \texttt{focus}; light gray: \texttt{focus\_prewhiten}.}
    \label{fig:multiple para test order p unknown larger data size(part3_2_add_p_known)}
\end{figure}

We now finally combine order selection with parameter estimation by applying the AIC-based model selection procedure as mentioned in Section \ref{tuning threshold}. Figure \ref{fig:multiple para test order p unknown larger data size(part3_2_add_p_known)} repeats the experiments of Figure \ref{fig:probation_size_effect_ar3} under this more realistic setting. 
The curves with known true order are different across training sizes as estimation of AR parameters are possibly different even if orders are the same.
Interestingly, the results are strikingly similar to the setting where the true order was assumed known. This indicates that AIC-based order selection reliably identifies the correct order from the probation data, and that the overall performance of \texttt{AR($p$)-focus} is not meaningfully degraded by the additional uncertainty in order estimation, at least for the probation sizes considered here.

\section{Accounting for Temporal Dependence in Real-Time Monitoring of Network Traffic Data}\label{sec:real-application}

To further illustrate the performance of \texttt{AR($p$)-focus} we consider an application related to network traffic data provided by an industry partner with the aim of monitoring the performance of network devices. For testing and development purposes, three high-frequency datasets were collected, each one consisting of approximately $800$,$000$ observations (we display in Figure \ref{fig:intro_data} a portion of the first dataset). To evaluate how an approach modelling for temporal dependence would behave on such data, we compare the same three approaches of the simulation study: (i) the proposed \texttt{AR($p$)-focus} method, (ii) the \texttt{focus} method assuming IID noise, and (iii) the \texttt{focus\_prewhiten} approach. Since the data contain multiple changepoints and collective anomalies (sometimes referred as epidemic changes), we use a simple sequential detection scheme: we process new data points and compute the statistics sequentially one observation at a time, and upon crossing the threshold, we record the change point and restart detection immediately. %

\begin{figure}
    \centering
    \includegraphics[width=0.8\linewidth]{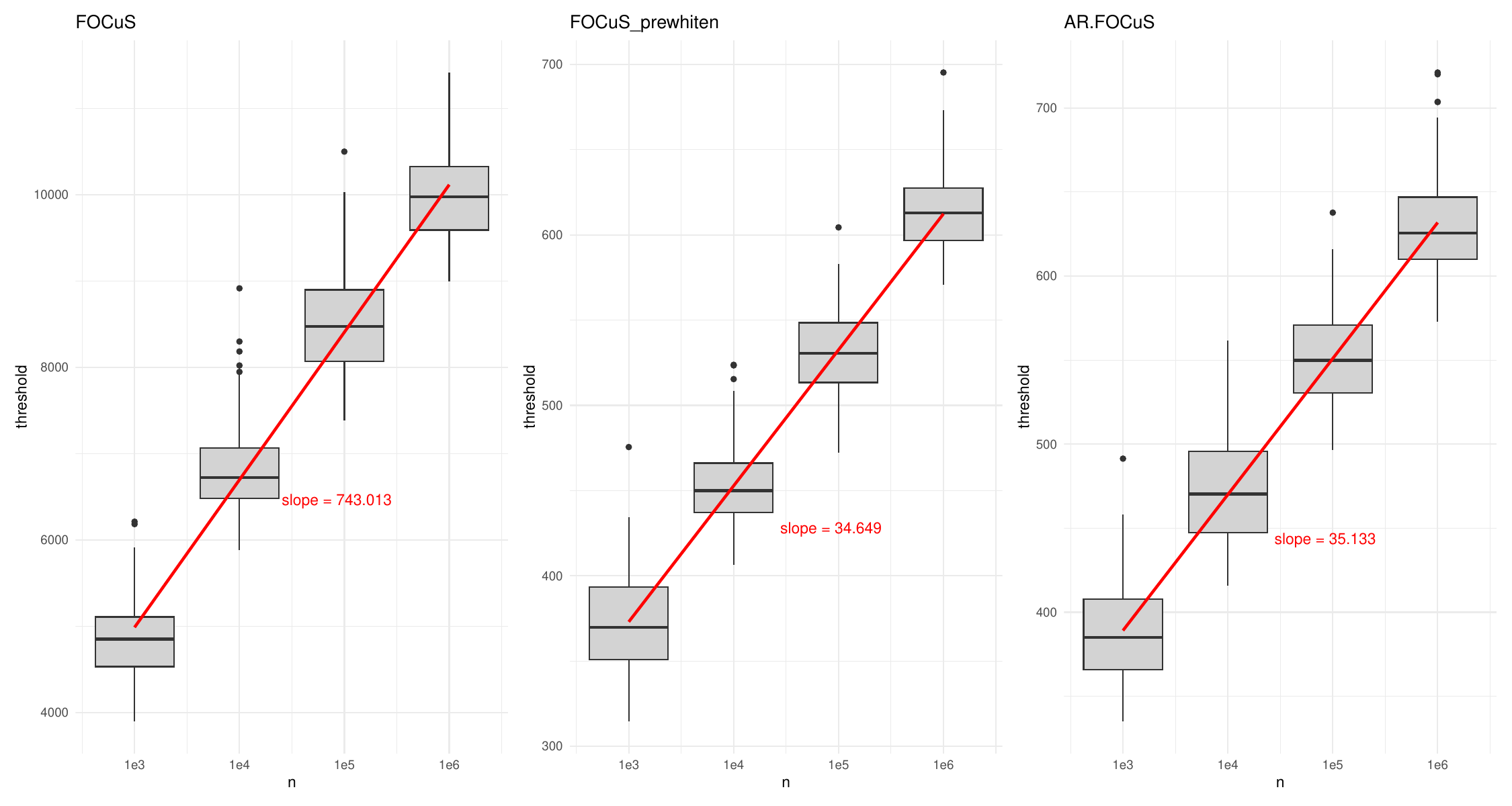}
    \caption{Calibrated thresholds as a function of stream length $n$ for \texttt{focus} (IID), \texttt{focus} (IID) on pre-whitened data, and \texttt{AR(p)-focus}, estimated via Monte Carlo simulation. The red line indicates the fitted linear trend on the $\log(n)$ scale, with the estimated slope as a label.}
    \label{fig:tresh-comparison}
\end{figure}

For each dataset, we estimate AR parameters using a probation period of 1000 observations, selecting the AR model order via the AIC, as in Section \ref{sec:sim_estimation}. Thresholds are then tuned using the Monte Carlo procedure described in Section~\ref{tuning threshold}. %
Figure \ref{fig:tresh-comparison} illustrates how the distribution of the maximum LR statistic under the null varies as a function of the stream length, $n$, across the three methods of this study. We denote how the statistics produced by the IID\ \texttt{focus} method are substantially inflated relative to those of the methods correcting for autocorrelation. To maintain a target false positive rate, the detector must adopt an artificially higher threshold then those accounting for correlation. By contrast, the thresholds for \texttt{focus\_prewhiten}, and \texttt{AR(p)-focus} are lower and not significantly different. For the rest of this section, all methods are calibrated to achieve a false positive rate of 0.01 over one million iterations under the null, as false detections are deemed costly as they require human intervention.

As we are interested in comparing performances of an IID\ detector with one that accounts for temporal correlation, we employ a comparative framework that avoids relying on ground truth knowledge of true changes. Specifically, we compare how quickly each method detects changes commonly identified all competing approaches.

Let $T_{i,j}$ denote the stopping time of the $i$the method over the $j$-th re-initialisation, defined as:
\begin{equation}
T_{i,j} = \inf\left\{t \geq T_{i,j-1} : \max_{\substack{\tau \in \{T_{i,j-1}+1, \dots, t\},\\  \mu_0, \mu_1}} 
\text{LR}^{*,T_{i,j-1}}_{\tau,t}(\mu_0, \mu_1) > \lambda^{(i)} \right\},
\end{equation}
where $\text{LR}^{*,T_{i,j-1}}_{\tau,t}$ is the likelihood ratio statistic computed at 
time $t$ based on data after $T_{i,j-1}$ with changepoint at $\tau \in [T_{i,j-1}, t]$, $\mu_0$ 
and $\mu_1$ are the pre- and post-change parameters, and $\lambda^{(i)}$ is the 
detection threshold for the $i$th method. The corresponding estimated changepoint is:
\begin{equation}
\hat{\tau}_{i,j} = \arg\max_{\substack{\tau \in \{T_{i,j-1}+1, \dots, T_{i,j}\},\\  \mu_0, \mu_1}} 
\text{LR}^{*,T_{i,j-1}}_{\tau,T_{i,j}}(\mu_0, \mu_1).
\end{equation}

\begin{figure}
    \centering
    \includegraphics[width=.9\textwidth]{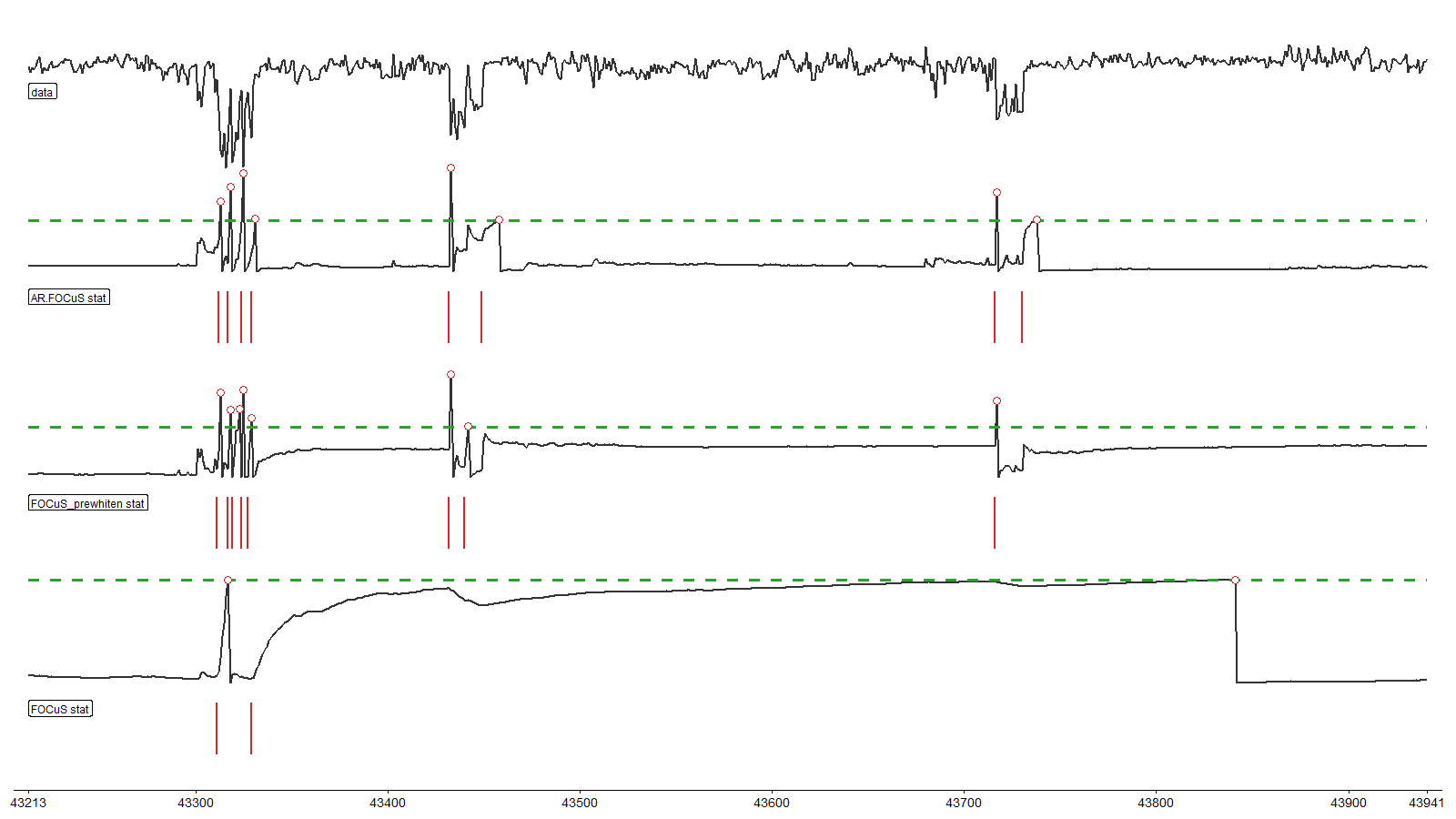}
    \caption{(Top) a subset of one of the analysed time series, with corresponding traces for both AR($p$)-detector (middle top), the IID detector on pre-withened data (middle bottom) and the regular IID detector (bottom). On each trace, the green dotted line represents threshold. The red dot indicates the stopping time: for each stopping time, the estimated change is illustrated by a red segment.}
    \label{fig:single cut of detection}
\end{figure}

Two methods are considered to detect the same change if their estimated changepoints lie within a tolerance of $\delta = 10$ observations, e.g. for methods $i$ and $k$, and respective changes $\hat{\tau}_{i,j}$, $\hat{\tau}_{k,\ell}$, if $|\hat{\tau}_{i,j} - \hat{\tau}_{k,\ell}| \leq \delta$. For each pair of methods detecting an agreed change, we define the true changepoint as $\hat{\tau}_j^* = \min(\hat{\tau}_{i,j}, \hat{\tau}_{k,\ell})$, and compute the detection delay for method $i$ as $T_{i,j} - \hat{\tau}_j^*$. An example of a subset of data together with the output of different methods and their estimated change-points is shown in Figure \ref{fig:single cut of detection}.

We report two metrics: (1) the total number of estimated changepoints identified by each method, in Table \ref{tab:focus_compare_table} (a), and (2) the pairwise average detection delay across all commonly detected changes, in Table \ref{tab:focus_compare_table} (b)--(c). This allows us to assess both detection sensitivity, and how quickly each method detects mutually agreed-upon changes.

We see that IID \texttt{focus} tends to detect far fewer changes, and on the commonly detected ones, it shows significantly slower detection delays on average. The reasons behind this disparity appear clear when we note that the IID detector consistently misses the second change in short, low magnitude collective anomalies. For example see Figure \ref{fig:single cut of detection}: we notice how, despite having correctly identified the beginning of a collective anomaly, the detector built on top of the IID \texttt{focus} misses all subsequent changes, with the statistics passing the change only after two further collective anomalies. As we highlighted in the simulation study, to control for the unaccounted temporal dependence there is an inherent loss of power, and this is mostly seen on hard-to-detect changes. Compared to the pre-whitened detector, which accounts for temporal dependence, while \texttt{focus\_prewhiten} improves detection of those smaller intensity collective anomalies, our proposed \texttt{AR($p$)-focus} still detects a greater number of changepoints overall, whilst achieving comparable or lower average detection delays across all three datasets on commonly detected changes, yielding an improvement in sensitivity.

\begin{table}[htbp]
\centering

\begin{subtable}{\textwidth}
\centering
\begin{tabular}{c rrr}
\toprule
Dataset & \texttt{focus} & \texttt{focus} (prewhiten) & \texttt{AR($p$)-focus} \\
\midrule
1 & 889  & 3100 & 4030 \\
2 & 813  & 2672 & 2697 \\
3 & 1340 & 2519 & 2890 \\
\bottomrule
\end{tabular}
\caption{}

\end{subtable}

\vspace{1em}

\begin{subtable}[t]{0.48\textwidth}
\centering
\begin{tabular}{c rr}
\toprule
Dataset & \texttt{focus} & \texttt{AR($p$)-focus} \\
\midrule
1 & 29.29 & 1.90 \\
2 &  5.82 & 1.44 \\
3 & 41.97 & 6.41 \\
\bottomrule
\end{tabular}
\caption{\texttt{focus} vs \texttt{AR($p$)-focus}.}

\end{subtable}
\hfill
\begin{subtable}[t]{0.48\textwidth}
\centering
\begin{tabular}{c rr}
\toprule
Dataset & \texttt{focus} (prewhiten) & \texttt{AR($p$)-focus} \\
\midrule
1 & 3.59 & 1.75 \\
2 & 2.56 & 2.59 \\
3 & 9.83 & 6.30 \\
\bottomrule
\end{tabular}
\caption{\texttt{focus} (prewhiten) vs \texttt{AR($p$)-focus}.}

\end{subtable}

\caption{Comparison of \texttt{focus} and \texttt{AR($p$)-focus} across the three datasets. 
(a) Total number of estimated changepoints detected by each method. 
(b)--(c) Average detection delay for commonly detected changepoints across methods.}
\label{tab:focus_compare_table}
\end{table}

\section{Discussion}
 
We have extended the \texttt{focus} algorithm for online Gaussian change-in-mean detection, originally developed under the assumption of IID\ noise, to the setting of AR($p$) noise, proposing \texttt{AR($p$)-focus} as a computationally efficient solution. Both simulation studies and real data applications confirm that explicitly modelling the autocorrelation structure yields meaningful reductions in detection delay relative to methods that ignore it or apply pre-whitening as a post-hoc correction, provided that sufficiently accurate estimates of the AR($p$) parameters are available. In practice these must be estimated from a probation period, which may appear a limitation: however, the settings in which \texttt{AR($p$)-focus} is most naturally applied are precisely those involving high-frequency data, where a sufficient probation record is easily available. As demonstrated in the real data application of Section \ref{sec:real-application}, where less then $0.2\%$ of the available data was used for probation, this translates directly into faster and more reliable detections.

The current framework is restricted to univariate time series, and extending it directly to the multivariate setting is non-trivial. A natural direction for future work is to build on \texttt{md-focus} from \cite{mdFOCuS}, which carries the \texttt{focus} framework to the multivariate case under IID\ noise, and to incorporate the AR($p$) modelling developed here into that richer setting. More broadly, relaxing the parametric AR($p$) assumption to accommodate longer-range dependence, or extending to non-stationary or seasonal signal structures would further widen the practical applicability of the approach, as for instance achieved by the approach \cite{romano2022detecting} in the offline case.

\begin{description}

\item[Section \ref{sec:sim_estimation} Replication Code:] The \texttt{AR($p$)-focus} method and simulation study described in the article is available in a pure R implementation at:\\ \url{ https://github.com/Yuntang2023/AR_FOCuS-simulation-code.git}.

\item[R and Python Packages:] A full implementation in C++  of \texttt{AR($p$)-focus} will be available in the near future in CRAN package \texttt{focus} (R interface) and on PyPI package \texttt{focus-py} (Python interface).

\end{description}

\bibliographystyle{apalike}
\bibliography{ref}

\newpage
\begin{center}
{\large\bf SUPPLEMENTARY MATERIAL}
\end{center}

\appendix

\input{appendix}

\end{document}

%% file: appendix.tex
\section{Proof of Proposition \ref{Prop:1}}\label{appendix proof}
We use notation $S_{t,\tau}(\mu_{1})$, $C_{k,t}(\mu_{1})$, $C_{t}(\mu_{1})$ defined in Section \ref{method of AR.FOCUS}. Given that $LR_n(\mu_1)$ is the likelihood-ratio statistic conditional on $\mu_1$ but where we have maximised over $\tau$, then we can write it as
\[
LR_n(\mu_1)=\max\left\{ ~~LR_{0,n}(\mu_{1}),\,LR_{1,n}(\mu_{1}),\,\ldots,\,LR_{n-1,n}(\mu_{1}) ~~
\right\}.
\]

For $n \leq p$, $LR_{\tau,n}(\mu_{1})$ is equal to the corresponding $S_{n,\tau}(\mu_{1})$ function.

For $n \geq (p+1)$,%
we define $Q_{n}(\mu_{1})$ as the likelihood-ratio statistic conditional on $\mu_1$ where we have maximised over a collection $n-\tau>p$:
\begin{align}\label{Q_n definition}
Q_{n}(\mu_{1}) &= \max\left\{ ~~LR_{0,n}(\mu_{1}),\,LR_{1,n}(\mu_{1}),\,\ldots,\,LR_{n-p-1,n}(\mu_{1}) ~~
\right\}, %
\end{align}
so that
\[
LR_n(\mu_1)=\max\left\{ ~~\max_{i=1,\ldots,p} \{S_{n,n-i}(\mu_1)\}~~ , ~~ Q_n(\mu_1) ~~ 
\right\}.
\] 

For $n \geq p+2$, we know $LR_{n-i,n}(\mu_{1})=LR_{n-i,n-1}(\mu_{1})+C_{n}(\mu_{1})$ for $i>p$. So, by definition of $Q_{n}(\mu_{1})$ in \ref{Q_n definition}, we get
\begin{align*}
    Q_{n}(\mu_{1}) &= \max_{i=(p+1),\ldots,n} \{LR_{n-i,n-1}(\mu_1)\}~~ + C_{n}(\mu_{1}) \\
&= \max\left\{ ~~LR_{(n-p-1),n-1}(\mu_{1})~~,~~\max_{i=(p+1),\ldots,(n-1)} \{LR_{(n-1)-i,n-1}(\mu_1)\}~~ 
\right\} + C_{n}(\mu_{1}) \\
&= \max\bigg\{~~
S_{n-1,n-p-1}(\mu_1)~~,~~Q_{n-1}(\mu_1)~~\bigg\} + C_{n}(\mu_1).
\end{align*} 

For $n=p+1$, in this case $Q_{p+1}(\mu_1) = LR_{0,p+1}(\mu_{1})$, and by definition $LR_{0,p+1}(\mu_{1}) = S_{p,0}(\mu_1)+C_{p+1}(\mu_1)$, which gives the initial condition for $Q_{n}(\mu_{1})$.

\section{Proof of Proposition \ref{Prop:unknown mean}} \label{Appendix: proof unknown mean}

The log-likelihoods $\ell_{\tau,n}(\mu_0,\mu_1)$  and $\ell_{\tau',n}(\mu_0,\mu_1)$ only differ in terms of the contribution from observations at times $t=\tau+1,\ldots,\tau'+p$, and this gives
\begin{eqnarray*}
2\lefteqn{[ \ell_{\tau,n}(\mu_0,\mu_1)-\ell_{\tau',n}(\mu_0,\mu_1)] = }  \\  & &
-\sum_{t=\tau+1}^{\tau'+p}  (y_t-v_{t-\tau}(\mu_1-\mu_0)-v_{p+1}\mu_0)^2 +\sum_{t=\tau+1}^{\tau'} (y_t-v_{p+1}\mu_0)^2 
+\sum_{t=\tau'+1}^{\tau'+p} (y_t-v_{t-\tau'}(\mu_1-\mu_0)-v_{p+1}\mu_0)^2.
\end{eqnarray*}
Writing $\Delta=\mu_1-\mu_0$ and $\mu=v_{p+1}\mu_0$ and rearranging gives
\begin{eqnarray*}
\lefteqn{ 2\left[\ell_{\tau,n}(\mu_0,\mu_1)-\ell_{\tau',n}(\mu_0,\mu_1)\right]  }  \\  & = &
\sum_{t=\tau+1}^{\tau'+p} \left\{ 2v_{t-\tau} \Delta(y_t-\mu) - v_{t-\tau}^2\Delta^2 \right\}
- \sum_{t=\tau'+1}^{\tau'+p} \left\{2\Delta v_{t-\tau'}(y_t-\mu) - v_{t-\tau'}^2\Delta^2\right\}\\
&=& 
2\Delta \left(\sum_{t=\tau+1}^{\tau'+p} v_{t-\tau}(y_t-\mu) - \sum_{t=\tau'+1}^{\tau'+p}v_{t-\tau'}(y_t-\mu)\right)
-\Delta^2 \left(\sum_{t=\tau+1}^{\tau'+p} v_{t-\tau}^2 - \sum_{t=\tau'+1}^{\tau'+p}v_{t-\tau'}^2  \right) . 
\end{eqnarray*}
Now let, $h=\tau'-\tau$. Note that
\[
2\left(\sum_{t=\tau+1}^{\tau'} v_{t-\tau}+ \sum_{t=\tau'+1}^{\tau'+p}(v_{t-\tau}-v_{t-\tau'})\right)=2\left(\sum_{i=1}^{h+p}v_{i} -\sum_{i=1}^pv_i\right)=2hv_{p+1},
\]
and similarly
\[
 \sum_{t=\tau+1}^{\tau'} v_{t-\tau}^2 - \sum_{t=\tau'+1}^{\tau'+p}v_{t-\tau'}^2=
 \sum_{i=1}^{h+p} v_i^2-\sum_{i=1}^p v_i^2 = hv_{p+1}^2.
\]
Define
\[
B_{\tau,\tau'} = 2\left(\sum_{t=\tau+1}^{\tau'+p} v_{t-\tau}y_t  -\sum_{t=\tau'+1}^{\tau'+p}v_{t-\tau'}y_t\right),
\]
then, using $h=v_{p+1}\mu_0$
\[
2\left[ \ell_{\tau,n}(\mu_0,\mu_1)-\ell_{\tau',n}(\mu_0,\mu_1)\right] =  
-\Delta^2 hv_{p+1}^2 +\Delta(B_{\tau,\tau'}-2hv_{p+1}^2\mu_0). 
\]
The roots of the quadratic in $\Delta$ on the right-hand side are $\Delta=0$ and 
\[
\Delta=(B_{\tau,\tau'}-2hv_{p+1}^2\mu_0)/(hv_{p+1}^2)=\frac{B_{\tau,\tau'}}{hv_{p+1}^2}-2\mu_0.
\]
Thus, as the co-efficient of $\Delta^2$ is negative, this quadratic is positive for $\Delta$ between these two roots. This gives the required set of $(\mu_0,\mu_1)$ values in the proposition with $b_{\tau,\tau'}=B_{\tau,\tau'}/C_h$. For example, for positive change $\Delta>0$ and using $\Delta=\mu_1-\mu_0$, the range of $\mu_1$ values for which the loglikelihood for a change at $\tau$ is greater than a later change at $\tau'$ is $\mu_1>\mu_0$ and
\begin{equation} \label{eq:intercept}
\mu_1-\mu_0 < \frac{B_{\tau,\tau'}}{hv_{p+1}^2}-2\mu_0 \Rightarrow \mu_1 < \frac{B_{\tau,\tau'}}{hv_{p+1}^2}-\mu_0
 .
\end{equation}

\section{Pseudo-code for \texttt{AR($p$)-focus} with unknown pre-change mean}\label{pseudo-code unknown pre prune}
\SetKwComment{Comment}{/* }{ */}
\begin{algorithm}
\caption{\texttt{AR($p$)-focus} (one iteration)}\label{alg:1}
\KwData{$y_{n-p:n}$; $Q_{n-1}(\mu_{0}, \mu_{1})$ the cost function from the previous iteration.}
\KwIn{Detection threshold, $\lambda>0$, $\mu_{0}=0$}
Calculate $LR_{n-p-1,n-1}(\mu_{0}, \mu_{1})$ \Comment{See (\ref{coefficients unknown pre (n<2p)}) in Appendix \ref{coef updateing unknown pre}}
$Q_{n}(\mu_{0}, \mu_{1}) \gets \max{\Bigl\{LR_{n-p-1,n-1}(\mu_{0}, \mu_{1}), Q_{n-1}(\mu_{0}, \mu_{1})\Bigl\}}+C_{p+1,n}(\mu_{0}, \mu_{1})$ \Comment*[r]{Algorithm 2}
$\mathcal{Q}^{(1)}_{n} \gets \max_{\mu_{0}, \mu_{1}}{Q_{n}(\mu_{0}, \mu_{1})}$
\,;
~\\
$\mathcal{Q}^{(2)}_{n} \gets \max_{\mu_{0},\mu_{1},1 \leq i \leq p}{LR_{n-i,n}(\mu_{0}, \mu_{1})}$ \Comment*[r]{See (\ref{coefficients unknown pre (n<2p)}) in Appendix \ref{coef updateing unknown pre}
}
$\mathcal{Q}_{n} \gets \max\{\mathcal{Q}_{n}^{(1)},\,\mathcal{Q}_{n}^{(2)}\}$;
~\\
\If{$\mathcal{Q}_{n} \geq \lambda$}{
    \Return \emph{n as a stopping point}
}
\Return $Q_{n}(\mu_{0}, \mu_{1})$ \emph{for the next iteration}\,.
\end{algorithm}

\SetKwComment{Comment}{/* }{ */}
\begin{algorithm}
\caption{Algorithm for $\max{\Bigl\{LR_{n-p-1,n-1}(\mu_{0}, \mu_{1}), Q_{n-1}(\mu_{0}, \mu_{1})\Bigl\}}+C_{p+1,n}(\mu_{0}, \mu_{1})$ for $\mu_{1}>\mu_{0}$}\label{alg:2}
\KwData{$Q_{n}(\mu_{0}, \mu_{1})=Q$ an ordered set of tuples $\{q_{i} = (\tau_{i}, \mathbf{Coef_{i}}, l_{i})\,\, i = 1,\cdots,k\}$, $y_{n}$ }

\vspace*{1.0em}

Initialize $\mathbf{Coef}_{k+1}$ \Comment*[r]{See (\ref{coefficients unknown pre (n<2p)}) in Appendix \ref{coef updateing unknown pre}}

$q_{k+1} \gets (\tau_{k+1}=n-p-1,\,\,\mathbf{Coef}_{k+1},\,\,\l_{k+1}=\infty)$
\,;

$i \gets k$\,;
~\\
\While{$\operatorname{Inter}\!\Bigl(LR_{\tau_{k+1},n}(0,\mu_1),\,LR_{\tau_i,n}(0,\mu_1)\Bigl) \leq l_{i}$ \textbf{and} $i \geq 1$}{
    $i \gets i-1$\,;
}
$l_{k+1} \gets \operatorname{Inter}\!\Bigl(LR_{\tau_{k+1},n}(0,,\mu_1),\,LR_{\tau_i,n}(0,\mu_1)\Bigl)$\,;
~\\
\If{$i \neq k$}{
    $Q \gets Q\setminus \{q_{i+1}, \cdots, q_{k}\}$\,;
}
Update $\mathbf{Coef}_{j}$ for $j = 1,\cdots,i,k+1$ to account for $y_n$ \Comment*[r]{See (\ref{coefficients unknown pre (n>=2p)}) in Appendix \ref{coef updateing unknown pre}}

\vspace*{1.0em}

\Return $\{q_1,\ldots,q_i,\,\,q_{k+1}\}$.
\end{algorithm}

A description of \texttt{AR($p$)-focus} with unknown pre-change mean is given in Algorithm \ref{alg:1}. The algorithm consists of two main steps: (i) we solve the recursion in Proposition \ref{Prop:1} to obtain the function $Q_{n}(\mu_{0},\mu_{1})$ from $Q_{n-1}(\mu_{0},\mu_{1})$, and (ii) we maximise the function $Q_{n}(\mu_{0},\mu_{1})$ and $LR_{n-i,n}(\mu_{0},\mu_{1})$ for $1 \leq i \leq p$, then choose max value among them; we explain these two briefly as follows. Step 2, is the maximisation step, which trivially evaluates the Likelihood ratio test from either Equation \ref{eq:LR-known} or \ref{eq:LR-unknown} for each candidate quadratic stored in $\mathcal{Q}_{n}$, and picks the maximum overall. Step 1 is where the pruning is performed, and for a positive change $\mu_1>\mu0$ this is detailed in Algorithm \ref{alg:2}. 

In Algorithm \ref{alg:2} we store each $q_i$ an un-pruned change-point, $\tau_i$, the coefficients of the quadratic in $\mu_0$ and $\mu_1$ for the test-statistic for a change at $\tau_i$, and a value $l_i$. These sets are stored in the order such that smaller $i$ correspond to earlier changes, and the value $l_i$ indicates that a change at $\tau_i$ has higher likelihood than a change at $\tau_{i-1}$ for $\mu_1>l_i-\mu_0$.

We define $\operatorname{Inter}\!\left(LR_{\tau',n}(0,\mu_1),\,LR_{\tau,n}(0,\mu_1)\right)$ to be the non-zero value of $\mu_1$ for which
$$
LR_{\tau',n}(0,\mu_1) - LR_{\tau,n}(0,\mu_1)=0.
$$
If $\tau'>\tau$ then this is calculating the value $B_{\tau,\tau'}/(hv^2_{p+1})$, where $h=\tau'-\tau$, that appears in (\ref{eq:intercept}). This value shows, relative to $\mu_0$, the values below which the older change-point at $\tau$ has greater likelihood than the newer change at $\tau'$. 
Thus if we calculate 
$c=\operatorname{Inter}\!\left(LR_{\tau_{k+1},n}(0,\mu_1),\,LR_{\tau_i,n}(0,\mu_1)\right)$ then we have that a change at $\tau_{k+1}$ has higher likelihood than a change at $\tau_i$ for $\mu_1>c-\mu_0$, but $l_i$ is defined so that a change at $\tau_{i-1}$ has higher likelihood than a change at $\tau_i$ for $\mu_1<l_i-\mu_0$. Thus if $c<l_i$ then there are no values of $\mu_1$ for which the change at $\tau_i$ has higher likelihood than both the other two change locations, and it can be pruned. We repeat this check for pruning until a change-point cannot be removed.

If we consider a negative change size where $\mu_{1}<\mu_{0}$, then the initial condition $l_{k+1}=\infty$ in line $2$ is changed to $l_{k+1}=-\infty$ and the pruning criterion in line $5$ is changed to $$
\operatorname{Inter}\!\Bigl(LR_{k+1,n}(\mu_1),\,LR_{i,n}(\mu_1)\Bigl) \geq l_{i}.
$$

\section{Coefficients Update for the \texttt{AR($p$)-focus} Statistic}\label{coef updateing unknown pre}
The log-likelihood ratio statistics is the twice difference between log-likelihood under alternative and null, which we define as $Cost_{\tau,n}(\mu_{0}, \mu_{1})$ and $Cost_{\tau,n}(\mu_{0}, \mu_{0})$ respectively. We define $y_{t}=x_{t}-\rho_{1}x_{t-1}-\rho_{2}x_{t-2}-\cdots-\rho_{p}x_{t-p}$, $M_{n}=\sum_{t=1}^{n}y_{t}$.

Then for changepoint $\tau$ at current time point $n$,
\begin{align}\label{log-likelihood AR(p) unknown}
LR^*_{\tau,n}(\mu_{0}, \mu_{1}) = 2\biggl(Cost_{\tau,n}(\mu_{0},\mu_{1})-Cost_{\tau,n}(\mu_{0},\mu_{0})\biggl),
\end{align}
where
\begin{align*}
Cost_{\tau,n}(\mu_{0},\mu_{0}) = &-\frac{n}{2}v_{p+1}^{2}\mu_{0}^{2} + v_{p+1}M_{n}\mu_{0} - \frac{1}{2}\sum_{t=1}^{n}y_{t}^{2}.
\end{align*}
There are two cases of $Cost_{\tau,n}(\mu_{0},\mu_{1})$. One is for $n\leq\tau+p$,
\begin{align*}
Cost_{\tau,n}(\mu_{0}, \mu_{1}) 
= -\frac{1}{2} \Bigg(\ & \sum_{t=1}^{\tau}(y_{t}-v_{p+1}\mu_{0})^{2} + \sum_{t=\tau+1}^{n}(y_{t}-v_{t-\tau}\mu_{1}+u_{t-\tau}\mu_{0})^{2} \Bigg),
\end{align*}
the other one is for $n>\tau+p$,
\begin{align*}
Cost_{\tau,n}(\mu_{0}, \mu_{1}) 
= -\frac{1}{2} \Bigg(\ & \sum_{t=1}^{\tau}(y_{t}-v_{p+1}\mu_{0})^{2} \\
  & + \sum_{t=\tau+1}^{\tau+p}(y_{t}-v_{t-\tau}\mu_{1}+u_{t-\tau}\mu_{0})^{2} + \sum_{t=\tau+p+1}^{n}(y_{t}-v_{p+1}\mu_{1})^{2} \Bigg)
\end{align*}

The function $Cost_{\tau,n}(\mu_{0},\mu_{1})$ is a quadratic in $\mu_0$ and $\mu_1$ and can be written as $A_{\tau,n}\mu_{0}^{2}+B_{\tau,n}\mu_{1}^{2}+C_{\tau,n}\mu_{0}\mu_{1}+D_{\tau,n}\mu_{0}+E_{\tau,n}\mu_{1}+F_{\tau,n}$ for some constants $A_{\tau,n},\ldots,F_{\tau,n}$. Using the notation $v_i$, defined in Section \ref{method of AR.FOCUS}, the coefficients for case $n\leq\tau+p$ can be calculated as:

\begin{equation}
\begin{aligned}
A_{\tau,n} &= -\frac{1}{2}\biggl[
\tau v_{p+1}^{2}
+ \sum_{i=1}^{n-\tau}u_{i}^{2}
\biggr] \\
B_{\tau,n} &= -\frac{1}{2}
\sum_{i=1}^{n-\tau}v_{i}^{2}
\\
C_{\tau,n} &= \sum_{i=1}^{n-\tau}u_{i}v_{i} \\
D_{\tau,n} &= -\frac{1}{2}\biggl[
-2v_{p+1}M_{\tau}
+ 2\sum_{t=\tau+1}^{n}y_{t}u_{t-\tau}
\biggr] \\
E_{\tau,n} &= -\frac{1}{2}\biggl[
-2\sum_{t=\tau+1}^{n}y_{t}v_{t-\tau}
\biggr] \\
F_{\tau,n} &= -\frac{1}{2}\sum_{t=1}^{n}y_{t}^{2}
\end{aligned}
\label{coefficients unknown pre (n<2p)}
\end{equation}
If $n>\tau+p$, we found contributions adding on $Cost_{\tau,\tau+p}(\mu_{0},\mu_{1})$ depend on $y_{n}$ or $M_{n}$, the coefficients can be updated by changing $n$, $y_{n}$ or $M_{n}$, which is
\begin{equation}
\begin{aligned}
A_{\tau,n} &= A_{\tau,\tau+p} \\
B_{\tau,n} &= B_{\tau,\tau+p} - \frac{1}{2}(n-\tau-p)v_{p+1}^{2} \\
C_{\tau,n} &= C_{\tau,\tau+p} \\
D_{\tau,n} &= D_{\tau,\tau+p} \\
E_{\tau,n} &= E_{\tau,\tau+p}+v_{p+1}(M_{n}-M_{\tau+p}) \\
F_{\tau,n} &= F_{\tau,\tau+p} - \frac{1}{2}\sum_{t=\tau+p+1}^{n}y_{t}^{2}
\end{aligned}
\label{coefficients unknown pre (n>=2p)}
\end{equation}

\section{Additional Simulation Results}\label{appendix:simulations}

\subsection{Performance under the Known Pre-Change Mean}\label{appendix:known mean}

\begin{figure}[ht]
    \centering
    \begin{subfigure}{\textwidth}
        \centering
        \includegraphics[width=1.05\textwidth]{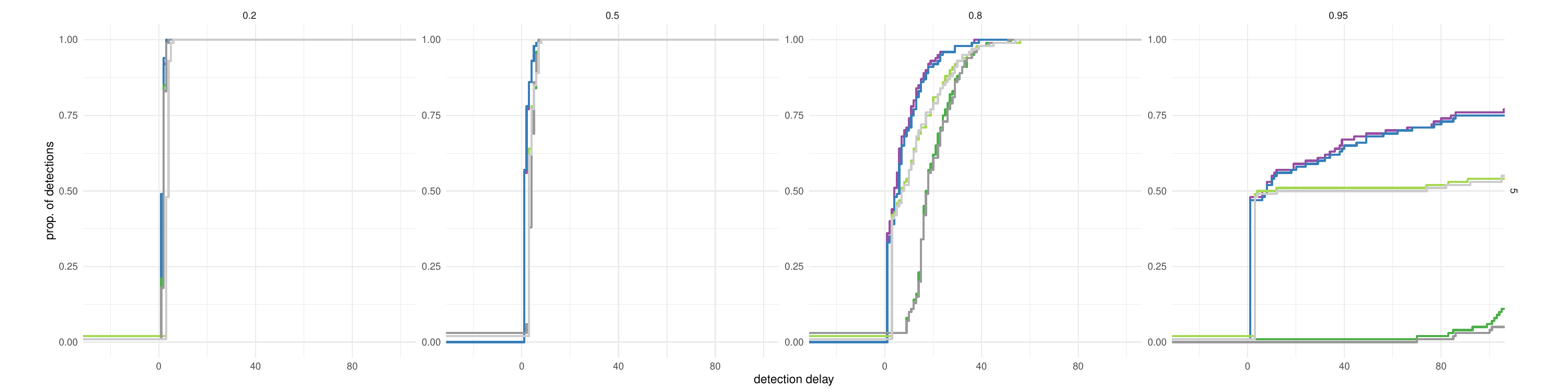}
        \caption{AR(1) noise processes}
        
    \end{subfigure}

    \vspace{1em}

    \begin{subfigure}{\textwidth}
        \centering
        \includegraphics[width=0.5\textwidth]{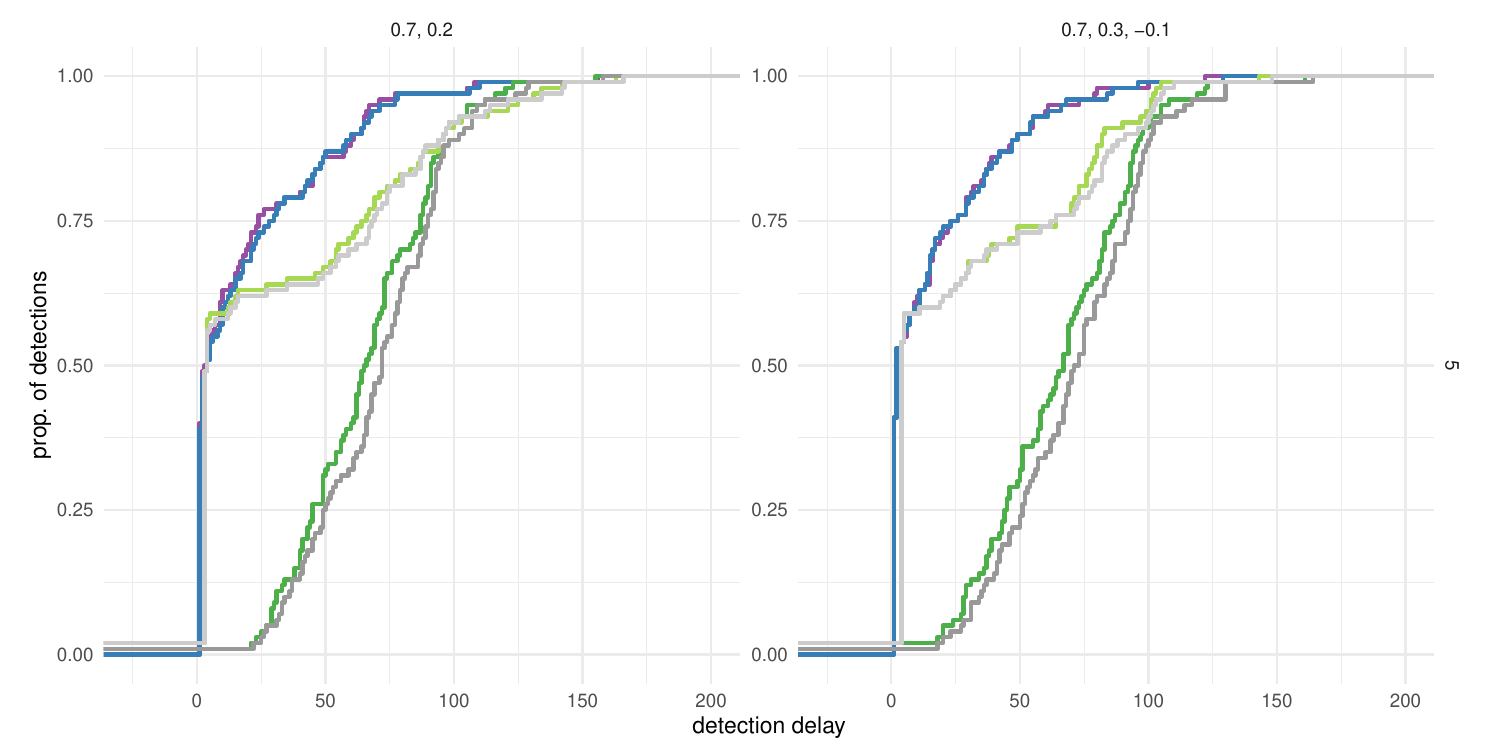}
        \caption{AR(2) and AR(3) noise processes}
        
    \end{subfigure}

    \caption{Oracle setting: empirical distribution of detection delay across 
    different noise processes, comparing known and unknown pre-change mean variants. 
    Data length $n = 10{,}000$, true changepoint at $5{,}000$, pre-change mean 
    $\mu_0 = 50$, change magnitude $\delta = 5$. Each column corresponds to a 
    different AR process. Purple: \texttt{AR($p$)-focus} (known mean, oracle); 
    blue: \texttt{AR($p$)-focus} (unknown mean, oracle); green: \texttt{focus} 
    (known mean, oracle); light green: \texttt{focus\_prewhiten} (known mean, 
    oracle); gray: \texttt{focus} (unknown mean, oracle); light gray: 
    \texttt{focus\_prewhiten} (unknown mean, oracle).}
    \label{fig:known_scenario}

\end{figure}

We complement the main simulation results by reporting performance in the known pre-change mean case, which serves as a further oracle benchmark. Figures \ref{fig:known_scenario} display the detection delay distributions for AR(1) and higher-order AR noise respectively, mirroring the structure of Figure \ref{fig:oracle_setting_comparison} in the main body, but now including both the known and unknown pre-change mean variants of each algorithm. All other simulation parameters take the same default values described in the main body.

As expected, algorithms with a known pre-change mean uniformly outperform their unknown-mean counterparts if training sets are sufficiently large, since they do not incur any estimation error for $\mu_0$. The relative ordering among methods otherwise mirrors the patterns described in Section~\ref{oracle setting performance compare}: differences are modest under weak autocorrelation and become more pronounced as the autocorrelation increases.

\begin{figure}[!ht]
    \centering
    \begin{subfigure}{\textwidth}
        \centering
        \includegraphics[width=\textwidth]{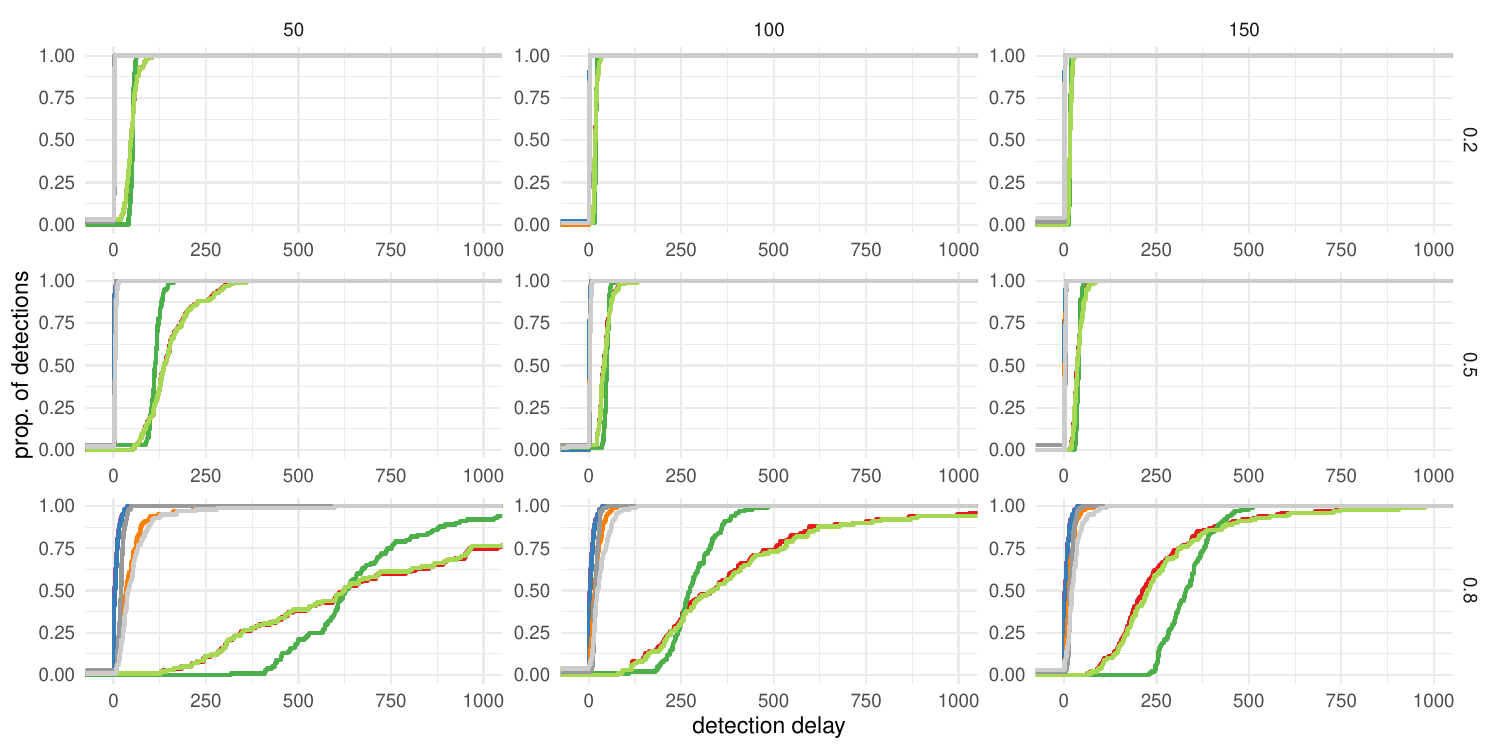}
        \caption{AR(1) noise processes}
        
    \end{subfigure}

    \vspace{1em}

    \begin{subfigure}{\textwidth}
        \centering
        \includegraphics[width=\textwidth]{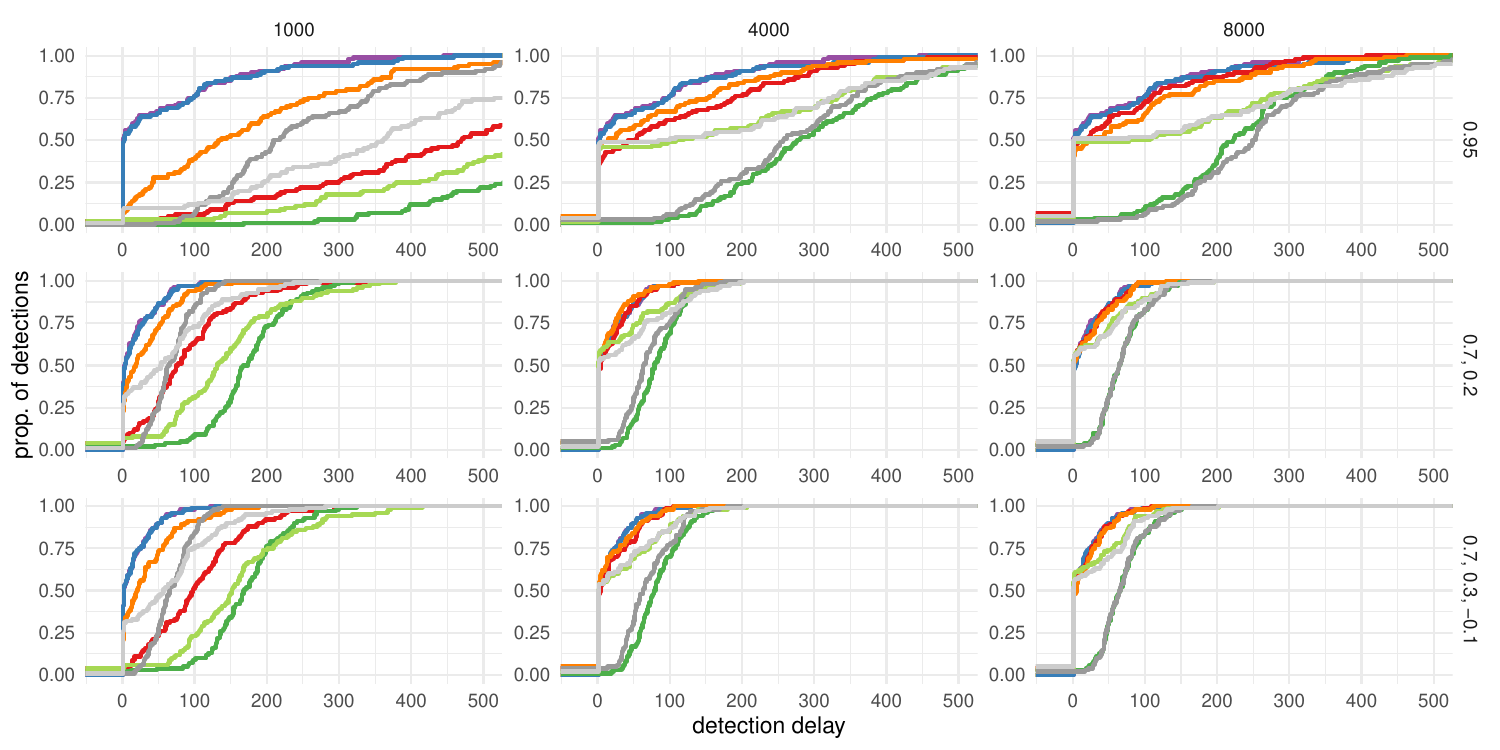}
        \caption{AR(1), AR(2), and AR(3) noise processes}
        \label{fig:probation_size_effect_ar3_full_results}
    \end{subfigure}

    \caption{Effect of probation period length on detection delay across different 
    noise processes. Each column corresponds to a different training size; each row 
    to a different AR process. Purple: \texttt{AR($p$)-focus} (known mean, oracle), blue: \texttt{AR($p$)-focus} (unknown mean, oracle); red: \texttt{AR($p$)-focus} (known mean); orange: \texttt{AR($p$)-focus} (unknown mean); gray: \texttt{focus}; light gray: \texttt{focus\_prewhiten}.}
    \label{fig:probation_size_effect_full_results}

\end{figure}

\begin{figure}[ht]
    \centering
    \includegraphics[width=1\textwidth]{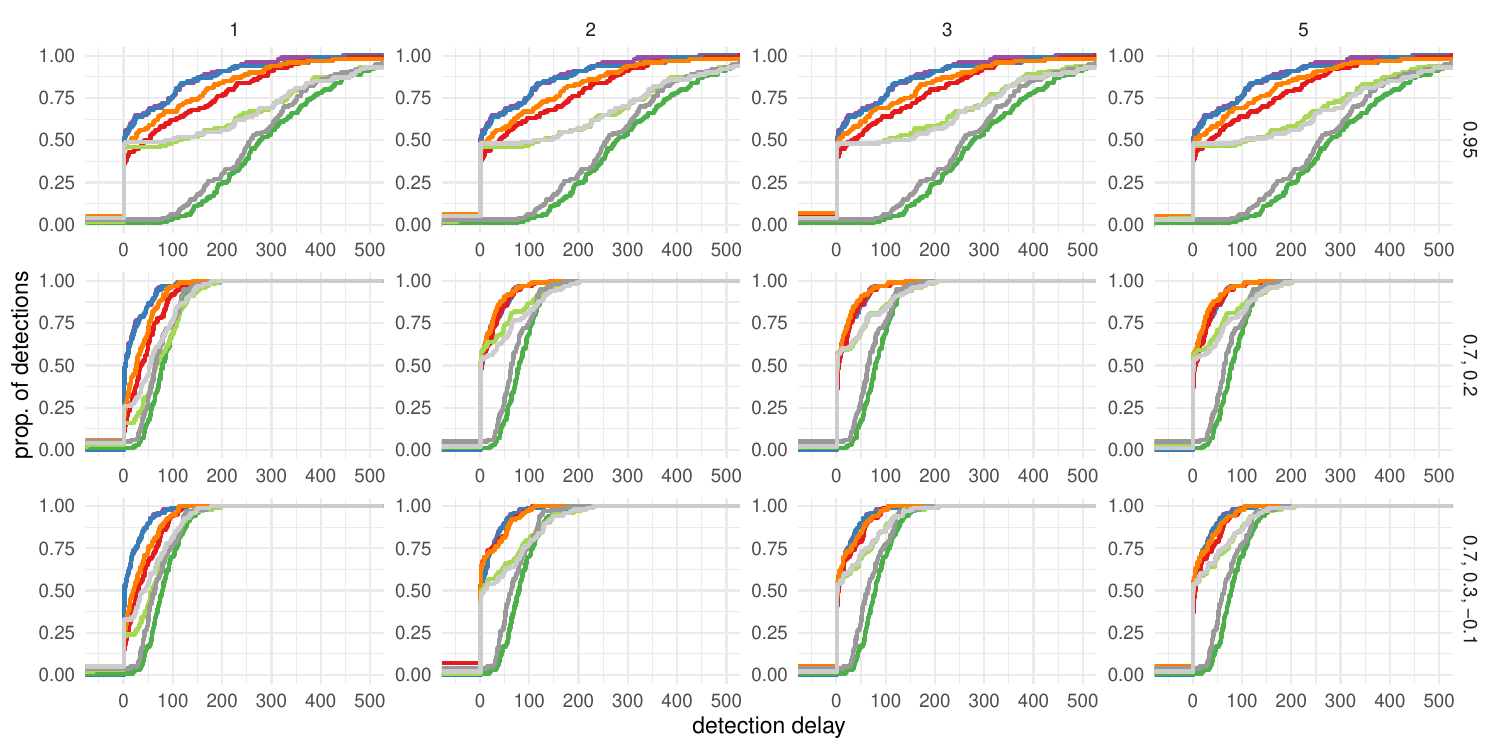}
    \caption{Robustness to AR order misspecification. Each row corresponds to a different true AR process (labelled by autocorrelation); each column to a different assumed order $p$ used by the algorithm. Purple: \texttt{AR($p$)-focus} (known mean, oracle), blue: \texttt{AR($p$)-focus} (unknown mean, oracle); red: \texttt{AR($p$)-focus} (known mean); orange: \texttt{AR($p$)-focus} (unknown mean); gray: \texttt{focus}; light gray: \texttt{focus\_prewhiten}.}
    \label{fig:Robustness order p test_full_results}
\end{figure}

\begin{figure}[ht]
    \centering
    \includegraphics[width=1\textwidth]{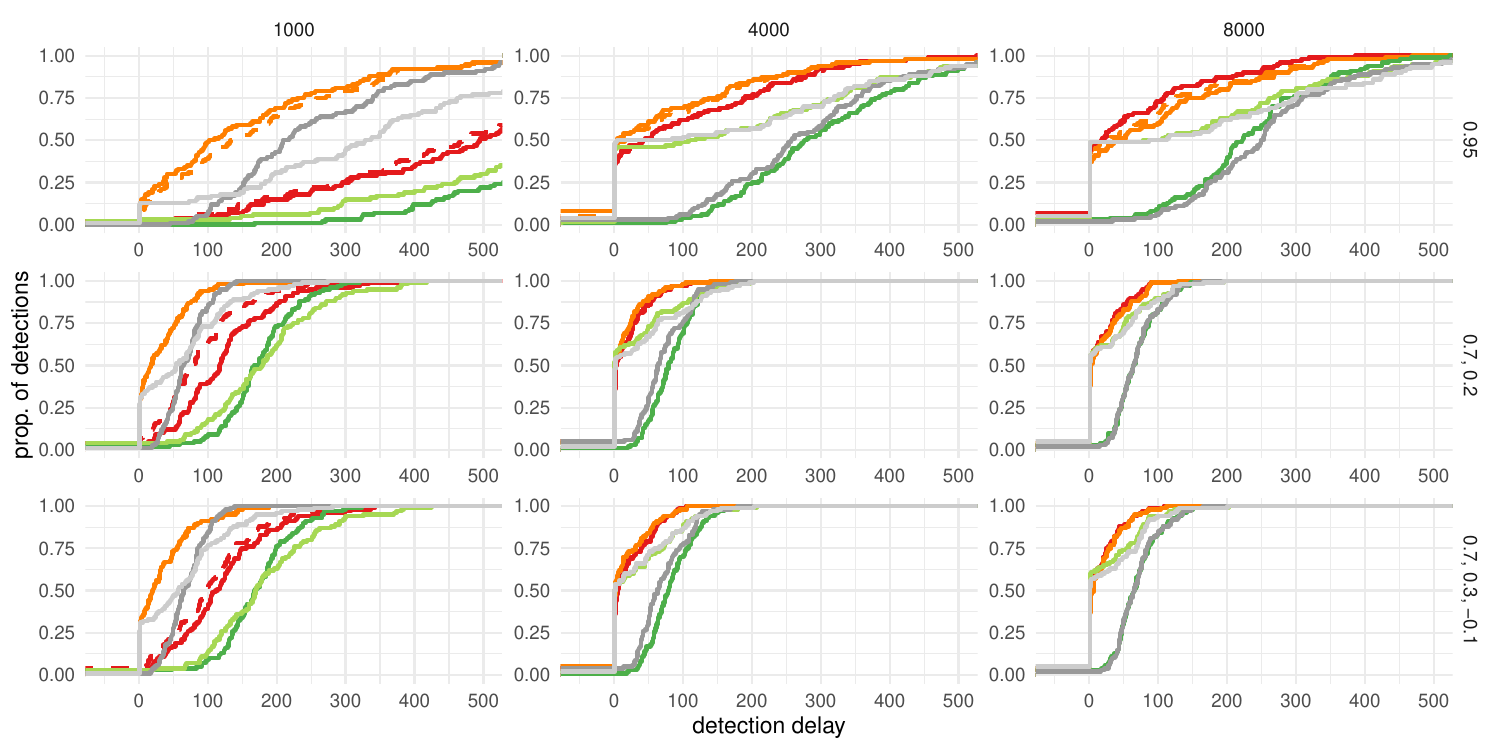}
    \caption{Performance of \texttt{AR($p$)-focus} when the AR order is unknown and selected by AIC. Each row corresponds to a different AR process (labelled by autocorrelation); each column to a different probation period length. Red dashed: \texttt{AR($p$)-focus} (known mean, p); red: \texttt{AR($p$)-focus} (known mean); orange dashed: \texttt{AR($p$)-focus} (known p); orange: \texttt{AR($p$)-focus}; gray: \texttt{focus}; light gray: \texttt{focus\_prewhiten}.}
    \label{fig:multiple para test order p unknown larger data size(part3_2_add_p_known) appendix}
\end{figure}

\begin{figure}[ht]
    \centering
    \includegraphics[width=1\textwidth]{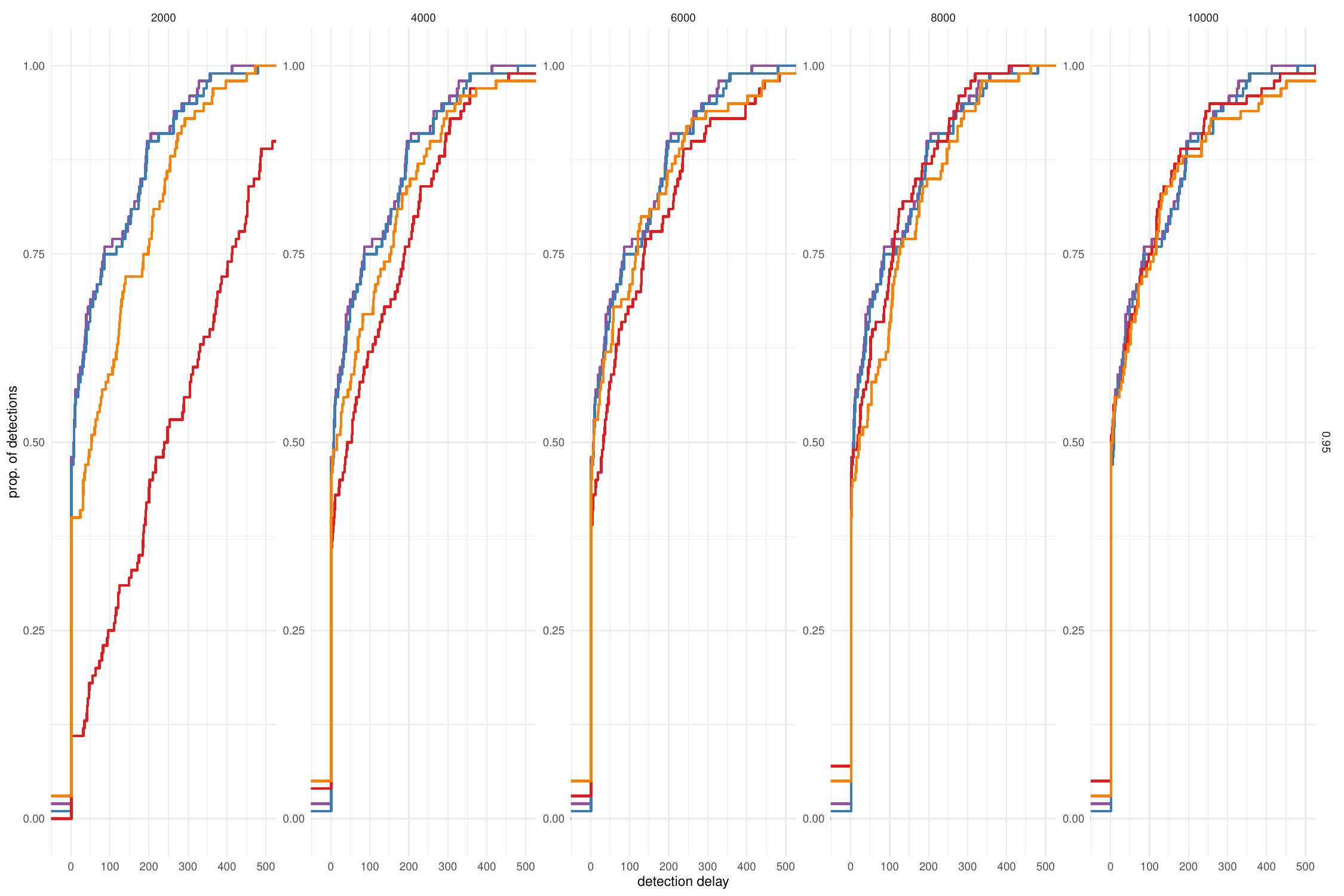}
    \caption{Effect of probation period length on detection delay: comparison between known and unknown pre-change mean variants of \texttt{AR($p$)-focus} for AR(1) noise. Each column corresponds to a different probetion period length. True changepoint at $5{,}000$, change magnitude $\delta = 5$, data length $n = 10{,}000$. Purple: \texttt{AR($p$)-focus} oracle (known mean); blue: \texttt{AR($p$)-focus} oracle (unknown mean); red: \texttt{AR($p$)-focus} (known mean); orange: \texttt{AR($p$)-focus} (unknown mean).}
    \label{fig:multiple para test (train size small to large)}
\end{figure}

We also examine how the relationship between the known and unknown pre-change mean cases evolves as the probation period length varies. Figure~\ref{fig:multiple para test (train size small to large)} compares \texttt{AR($p$)-focus} with its known-mean counterpart across a range of training sizes and AR(1) autocorrelation coefficients. When the size of training observations are shorter than true change $5000$, the unknown-mean algorithm is much better than known-mean performance as unknown-mean algorithm estimate the $\mu_{0}$ from pre-change observations. However, if the probation period is sufficiently long, the known-mean algorithm closely approaches and exceeds the unknown-mean benchmark.

\begin{figure}[ht]
    \centering
    \includegraphics[width=1\textwidth]{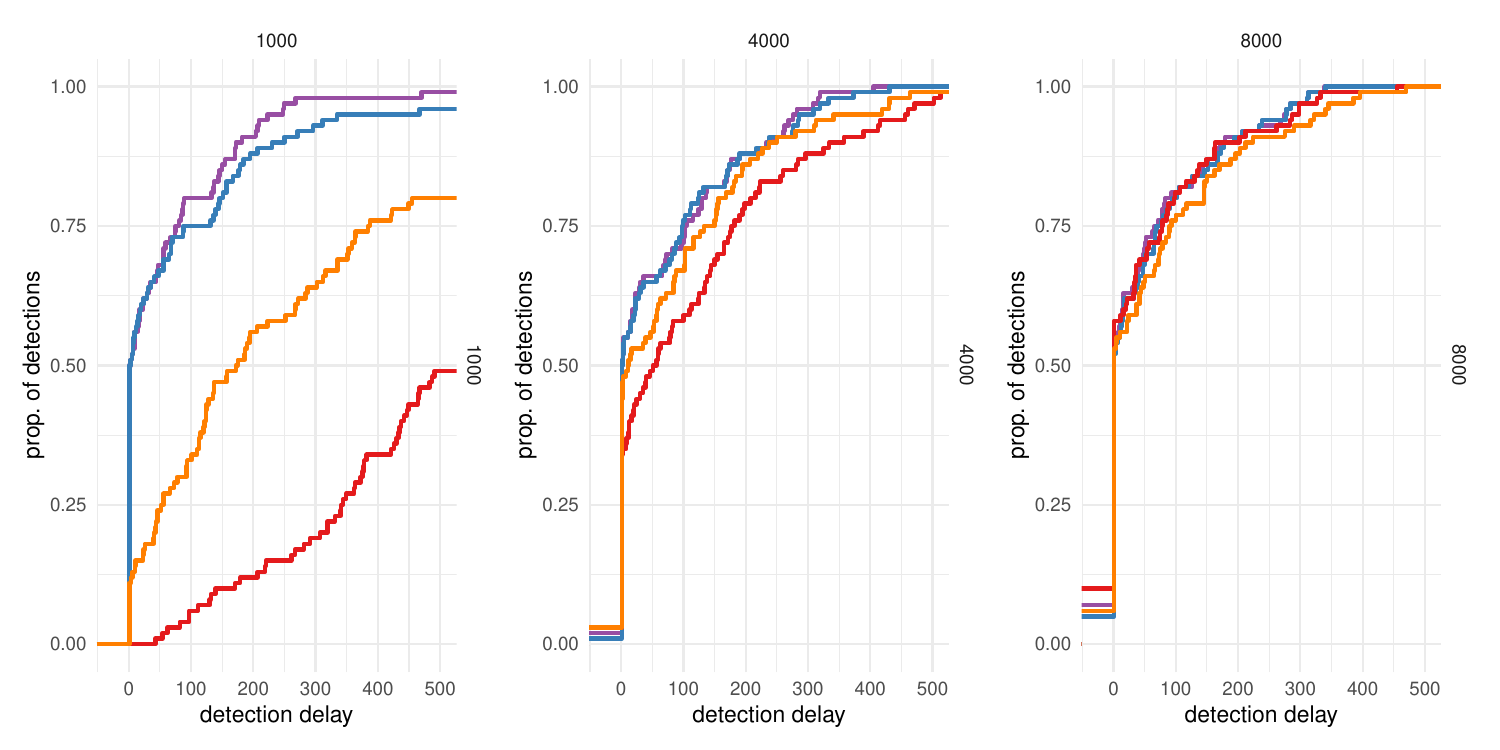}
    \caption{Comparison between known and unknown pre-change mean variants of \texttt{AR($p$)-focus} when the training size equals the true changepoint. The top and right margins of each panel indicate the training size and true changepoint, respectively. Change magnitude $\delta = 5$, data length $n = 10{,}000$. Purple: \texttt{AR($p$)-focus} oracle (known mean); blue: \texttt{AR($p$)-focus} oracle (unknown mean); red: \texttt{AR($p$)-focus} (known mean); orange: \texttt{AR($p$)-focus} (unknown mean).}
    \label{fig:multiple para test (combine_1)}
\end{figure}

To make a direct and fair comparison between the known and unknown pre-change mean cases, we set the training size equal to the true changepoint location, so that both algorithms have access to the same amount of pre-change data. Figure \ref{fig:multiple para test (combine_1)} shows results for a fixed data length of $n = 10{,}000$ as the training size and true changepoint are varied jointly. The known-mean variant dominates when the training sample is large, but its advantage erodes and reverse when the probation period is not very large, since the training set leads to poor estimation of the AR parameters regardless of whether the mean is known. 

\subsection{Effect of Maximum Assumed Order in the AR(1) Setting}\label{appendix:max order p}

To complement the misspecification analysis in Section~\ref{robustness order p}, we provide here a more detailed illustration restricted to the AR(1) case. This allows us to examine, in isolation, how the choice of maximum assumed order interacts with the strength of the AR(1) autocorrelation parameter.

\begin{figure}[ht]
    \centering
    \includegraphics[width=1\textwidth]{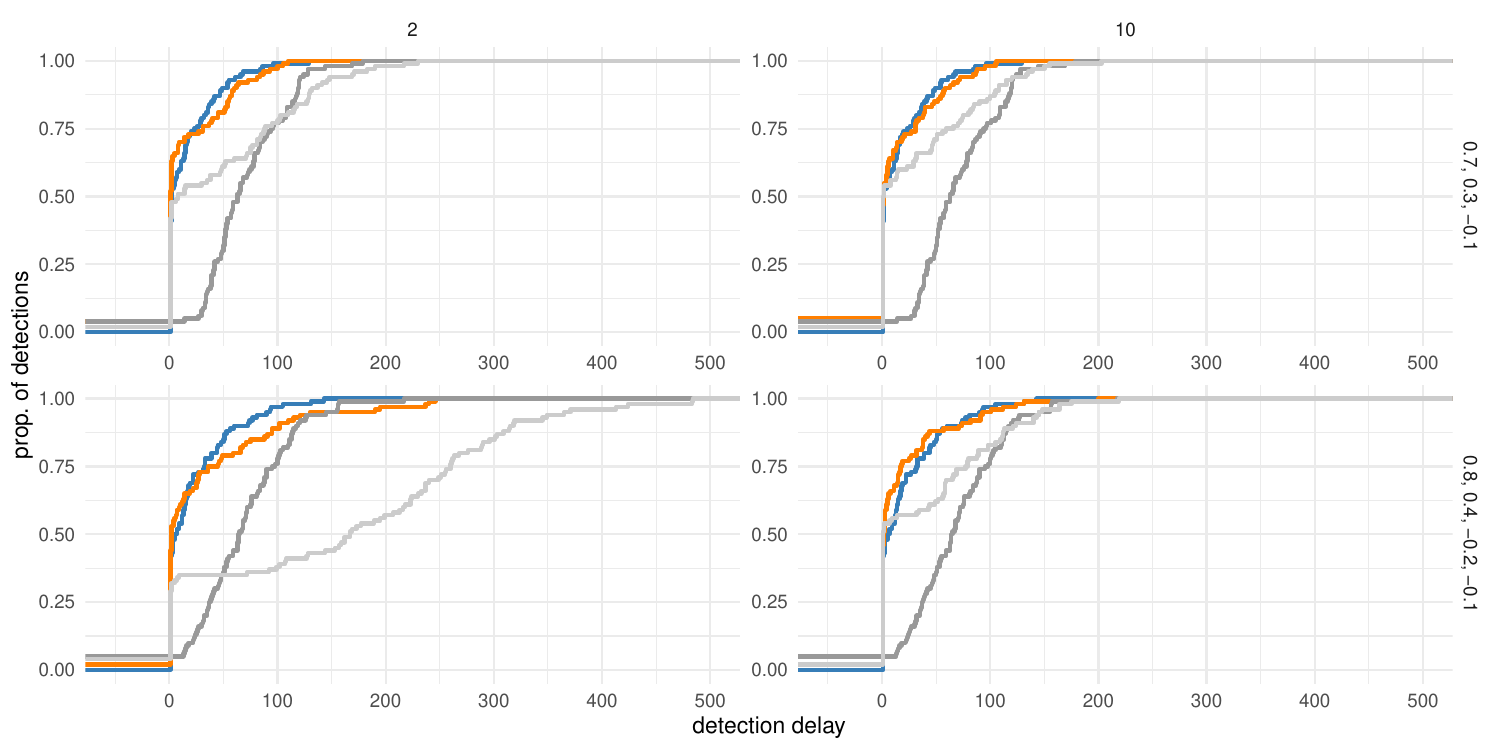}
    \caption{Effect of the maximum assumed AR order on detection performance in the AR(1) setting. Each row corresponds to a different AR(1) autocorrelation coefficient; each column to a different assumed maximum order. Blue: \texttt{AR($p$)-focus} oracle; orange: \texttt{AR($p$)-focus}; gray: \texttt{focus}; light gray: \texttt{focus}\_prewhiten.}
    \label{fig:max order p compare}
\end{figure}

Figure~\ref{fig:max order p compare} shows that assuming a maximum order larger than the true order carries little cost: since AIC will tend to select the correct (lower) order, or at worst fit a slightly over-parameterised model, the performance of \texttt{AR($p$)-focus} remains close to its oracle benchmark. By contrast, assuming a maximum order smaller than the true order forces the algorithm to work with a misspecified model that cannot account for all the relevant autocorrelation, resulting in a more substantial loss of power. This is consistent with the broader robustness results reported in Figure~\ref{fig:Robustness order p test}.

\subsection{Power Loss from Misspecification Error}\label{appendix:power loss iid}

The previous sections have focused on the gains from correctly modelling autocorrelation. Here we assess the consequences of imposing an AR model when the noise is in fact IID. This quantifies the penalty paid for model misspecification in the opposite direction, and is relevant whenever a practitioner applies \texttt{AR($p$)-focus} with a non-zero fixed AR parameter to data that does not actually exhibit autocorrelation.

To this end, we simulate data with IID noise and apply \texttt{AR($p$)-focus} with assumed AR(1) coefficients $\rho_1 \in \{0, 0.1, 0.3, 0.5, 0.7, 0.9\}$, for change magnitudes $\delta \in \{0.5, 1, 3, 5\}$. The assumed value $\rho_1 = 0$ corresponds to no misspecification and serves as the baseline.

\begin{figure}[ht]
    \centering
    \includegraphics[width=1\textwidth]{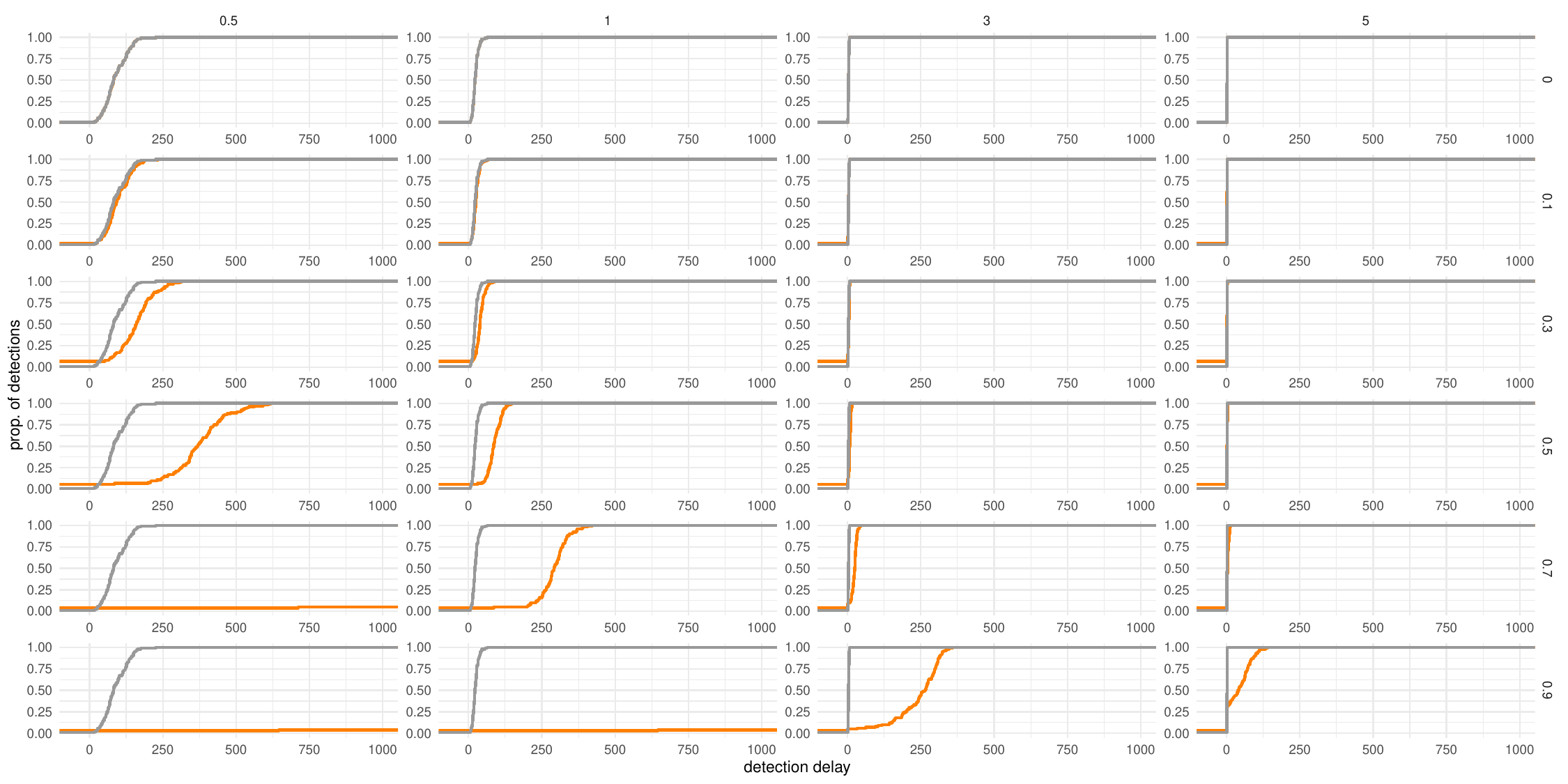}
    \caption{Power loss from misspecifying the AR parameter when the true noise is IID. Each column corresponds to a different change magnitude $\delta$; each row to a different assumed AR(1) coefficient $\rho_1$. Orange: \texttt{AR($p$)-focus}; gray: \texttt{focus}.}
    \label{fig:ar_assum_iid_mag}
\end{figure}

Figure~\ref{fig:ar_assum_iid_mag} shows that detection power decreases monotonically as the assumed AR parameter increases. The loss is most severe for small change magnitudes: at $\delta = 0.5$, the detection rate falls to near zero even for moderate values of $\rho_1$. For larger change magnitudes the method is considerably more robust; at $\delta = 5$, power remains high even when $\rho_1$ is assumed to be as large as $0.9$. These results underscore the importance of accurate AR parameter estimation when the change signal is weak: erroneous assumptions about the autocorrelation structure can render the detector ineffective, while the same misspecification is largely inconsequential when the signal is strong.

